\documentclass[manuscript,screen]{acmart}

\usepackage{multirow}
\usepackage{threeparttable}
\usepackage{breqn}
\usepackage{tikz}

\newcommand*\filledcircled[1]{\tikz[baseline=(char.base)]{
    \node[shape=circle, draw, inner sep=1pt, fill=black, text=white] (char) {#1};}}

\AtBeginDocument{%
  \providecommand\BibTeX{{%
    \normalfont B\kern-0.5em{\scshape i\kern-0.25em b}\kern-0.8em\TeX}}}

\begin{document}

\title{\textit{HEANA}: A Hybrid Time-Amplitude Analog Optical Accelerator with Flexible Dataflows for Energy-Efficient CNN Inference}

\author{Sairam Sri Vatsavai}
\email{ssr226@uky.edu}

\affiliation{%
  \institution{University of Kentucky}
  \country{USA}
}

\author{Venkata Sai Praneeth Karempudi}
\affiliation{%
  \institution{University of Kentucky}
  \country{USA}
}

\author{Ishan Thakkar}
\affiliation{%
  \institution{University of Kentucky}
  \country{USA}
}


\begin{abstract}
  Several photonic microring resonators (MRRs) based analog accelerators have been proposed to accelerate the inference of integer-quantized Convolutional Neural Networks (CNNs) with remarkably higher throughput and energy efficiency compared to their electronic counterparts. However, the existing analog photonic accelerators suffer from three shortcomings: (i) severe hampering of wavelength parallelism due to various crosstalk effects, (ii) inflexibility of supporting various dataflows with temporal accumulations, and (iii) failure in fully leveraging the ability of photodetectors to perform in-situ accumulations. These shortcomings collectively hamper the performance and energy efficiency of prior accelerators. To tackle these shortcomings, we present a  novel \underline{\textbf{H}}ybrid tim\underline{\textbf{E}}-\underline{\textbf{A}}mplitude a\underline{\textbf{N}}alog optical \underline{\textbf{A}}ccelerator, called HEANA. HEANA employs hybrid time-amplitude analog optical modulators (TAOMs) in a spectrally hitless arrangement which significantly reduces optical signal losses and crosstalk effects, thereby increasing the wavelength parallelism in HEANA. HEANA employs our invented balanced photo-charge accumulators (BPCAs) that enable buffer-less, in-situ, spatio-temporal accumulations to eliminate the need to use reduction networks in HEANA, relieving it from related latency and energy overheads. Moreover, TAOMs and BPCAs increase the flexibility of HEANA to efficiently support spatio-temporal accumulations for various dataflows. Our evaluation for the inference of four modern CNNs indicates that HEANA provides improvements of at least 25$\times$ and 32$\times$ in frames-per-second (FPS) and FPS/W (energy-efficiency), respectively, for equal-area comparisons, on gmean over two MRR-based analog CNN accelerators from prior work. 
\end{abstract}

%
%



\maketitle

\section{Introduction}
Deep Neural Networks (DNNs) achieve high inference accuracy, which has revolutionized their use in various artificial intelligence tasks, such as image recognition, language translation, and autonomous driving \cite{dnnapplications1,dnnapplications2}. Convolutional Neural Networks (CNNs) are specific types of DNNs \cite{cnnapplication}. CNNs are computationally intensive, and hence, require a long inference time. In CNNs, around 80\% of the total processing time is taken by convolution operations. The need to tackle the ever-increasing complexity and inference time of CNNs has pushed for highly customized CNN hardware accelerators \cite{Baischer2021}. For hardware acceleration of convolution operations, they can be converted into general matrix-matrix multiplication (GEMM) operations \cite{sze2017efficient}. The GEMM operations can be further decomposed into dot product operations to be efficiently mapped onto the hardware for acceleration. 

To effectively accelerate GEMM operations, it is crucial to support different dataflows like output stationary (OS), input stationary (IS), and weight stationary (WS), which dictate the order of computations and data movement. Dataflow selection can significantly impact the performance of a hardware accelerator \cite{maeri2018, kwon2019micro}. The optimal dataflow depends on factors such as matrix dimensions and convolution operations, which can vary across different layers within a CNN and different CNNs \cite{kwon2019micro}. Flexible dataflow support is necessary for efficient execution across diverse CNNs, enabling better runtime performance for various models. Additionally, for efficient and swift hardware-based acceleration, CNNs are quantized to have integer input/weight parameters \cite{krishnamoorthi2018quantizing}.


Among the CNN hardware accelerators from the literature, silicon-photonic accelerators have shown great promise to provide unparalleled parallelism, ultra-low latency, and high energy efficiency \cite{holylight,squeezelight,deapcnn,karen2020proceeding,crosslight}. Typically, a silicon-photonic CNN accelerator consists of multiple Dot Product Units (DPUs) that perform multiple dot product operations in parallel. Several DPU-based optical CNN accelerators have been proposed in prior works based on various silicon-photonic devices, such as Mach Zehnder Interferometer (MZI) (e.g., \cite{mzicomplex2021}, \cite{cansu2021}) and Microring Resonator (MRR) (e.g., \cite{deapcnn}, \cite{crosslight}).

Among these optical DPU-based CNN accelerators from prior work, the MRR-enabled analog DPU-based accelerators (e.g., \cite{holylight,deapcnn,squeezelight,pixel,crosslight,tait2017, sairamisqed2024}) have shown disruptive performance and energy efficiencies, due to the MRRs' compact footprint, low dynamic power consumption, and compatibility with cascaded dense-wavelength-division-multiplexing (DWDM). However, these accelerators face several challenges that hinder their scalability, throughput, and energy efficiency. These prior accelerators employ a combination of microring modulators (MRM) input array and MRR weight bank to perform dot product operation between input and weight values. The inter-modulation crosstalk in MRM input array and inter-spectral, electrical, and thermal crosstalk effects in MRR weight banks reduce the available optical power budget in DPUs. The reduction in the optical power budget significantly reduces the achievable DPU size and supported bit-precision \cite{lukasscalability,cases2022}. The MRR-based accelerators in prior works \cite{crosslight, deapcnn, cases2022} often focus on weight stationary dataflow only, as most of them use thermo-optic actuation of weights and high-speed electro-optic actuation of inputs. This results in a limited ability to support the IS and OS dataflows, as the weight values in these accelerators remain static (change less frequently) while the input and output values change more frequently. The IS and OS dataflows require frequent updating of weights, which prior accelerators cannot efficiently support due to the static nature of their weight banks. This constraint hinders the flexibility and efficiency of these accelerators for a variety of dataflows. Additionally, the IS and OS dataflows, along with the WS dataflow, generate partial sums that require frequent use of analog-to-digital converters and intermittent digital storage, which can lead to higher energy consumption and latency. Furthermore, the presence of crosstalk and high spectral sensitivity in MRR weight banks and MRM input arrays needs two separate feedback control units per MRM/MRR \cite{Tait18,Tait2022}, one for parameter (input/weight) tuning and one for thermal stability. This increases the static power consumption, diminishing the energy efficiency advantages. Moreover, none of the prior works have leveraged the ability of balanced photodetectors (BPDs) as in-situ spatio-temporal accumulators to enable flexible support for various dataflow and to reduce the required buffer accesses and corresponding latency and energy overheads.

 To address these shortcomings, this paper presents a novel \underline{\textbf{H}}ybrid \underline{\textbf{T}}ime-\underline{\textbf{A}}mplitude a\underline{\textbf{N}}alog optical \underline{\textbf{A}}ccelerator called HEANA. HEANA employs a novel design of hybrid Time-Amplitude Analog Optical Modulator (TAOM) in a spectrally hitless arrangement to eliminate spectral-interference and various crosstalk effects. Our TAOM uses a single active MRR to perform multiplication operations, thereby allowing electro-optic tuning of both input and weight values. Moreover, due to the single-MRR implementation of TAOMs, HEANA achieves a significant reduction in number of active MRR devices compared to prior works. This reduces not only the area consumption but also the insertion losses in HEANA, thereby increasing its energy efficiency. Moreover, HEANA introduces a novel approach to Balanced Photo-Charge Accumulators (BPCAs) that enhances the temporal accumulation of a high number of partial sums (psums) in situ. Unlike previous works \cite{alexandermit2022, BhaskaranPCA2022, oxbnn, sconna} that utilized balanced photodetectors and time-integrating receivers with single or alternating accumulation capacitors, HEANA employs multiple accumulation capacitors per time-integrating receiver to support multiple parallel accumulation frames. This design enables intermittent storage of multiple psums in the analog format directly on the accumulation capacitors, reducing the need for frequent analog-to-digital conversions and buffer accesses, and minimizing the footprint of psums in the utilized digital memory. These innovations allow HEANA to support different dataflows, including the weight, input, and output stationary dataflows. By enabling flexible support for multiple dataflows, HEANA achieves greater versatility in handling various computation tasks, significantly decreasing energy consumption and overall latency in CNN processing.


Our key contributions in this paper are summarized below.
\begin{itemize}
     \item We present our invented, novel, dataflow-flexible CNN accelerator called HEANA, which employs an array of hybrid time-amplitude analog optical modulators (TAOMs) in spectrally hitless DPU architecture and highly scalable in-situ spatio-temporal accumulators called Balanced Photo-Charge Accumulators (BPCAs).
     \item We present a novel approach using time-amplitude analog modulation in our TAOM, offering multifold benefits that include the reduction in the required dynamic range and signal power at the modulator. This leads to significant power savings. In addition, this approach requires only a single microring modulator (MRM) for multiplication, simplifying the design and implementation.
    \item We propose a temporal accumulation mechanism using BPCAs, which allows for efficient accumulation and storage of multiple partial sums in the analog format to enable efficient support for different dataflows such as the IS, WS, and OS dataflows.
    \item We perform detailed modeling and characterization of our invented TAOM and BPCA circuits using photonics foundry-validated, commercial-grade, photonic-electronic design automation tools (Section III).
    \item We propose a mapping strategy that supports the IS, OS, and WS dataflows while eliminating the need for external electronic partial sum reduction networks (Section IV).
    \item We perform a comprehensive scalability analysis of our HEANA DPUs, to determine their achievable maximum size (degree of achievable spatial parallelism) and supported bit-precision (Section V).
    
    \item We implement and evaluate HEANA for input stationary, output stationary, and weight stationary dataflows at the system-level. We compare its performance with two well-known photonic CNN accelerators from prior works for the inference of
four state-of-the-art CNNs with two different batch sizes (Section VI).
\end{itemize}

\begin{figure}[] 
    \centering
    \includegraphics[scale=0.32]{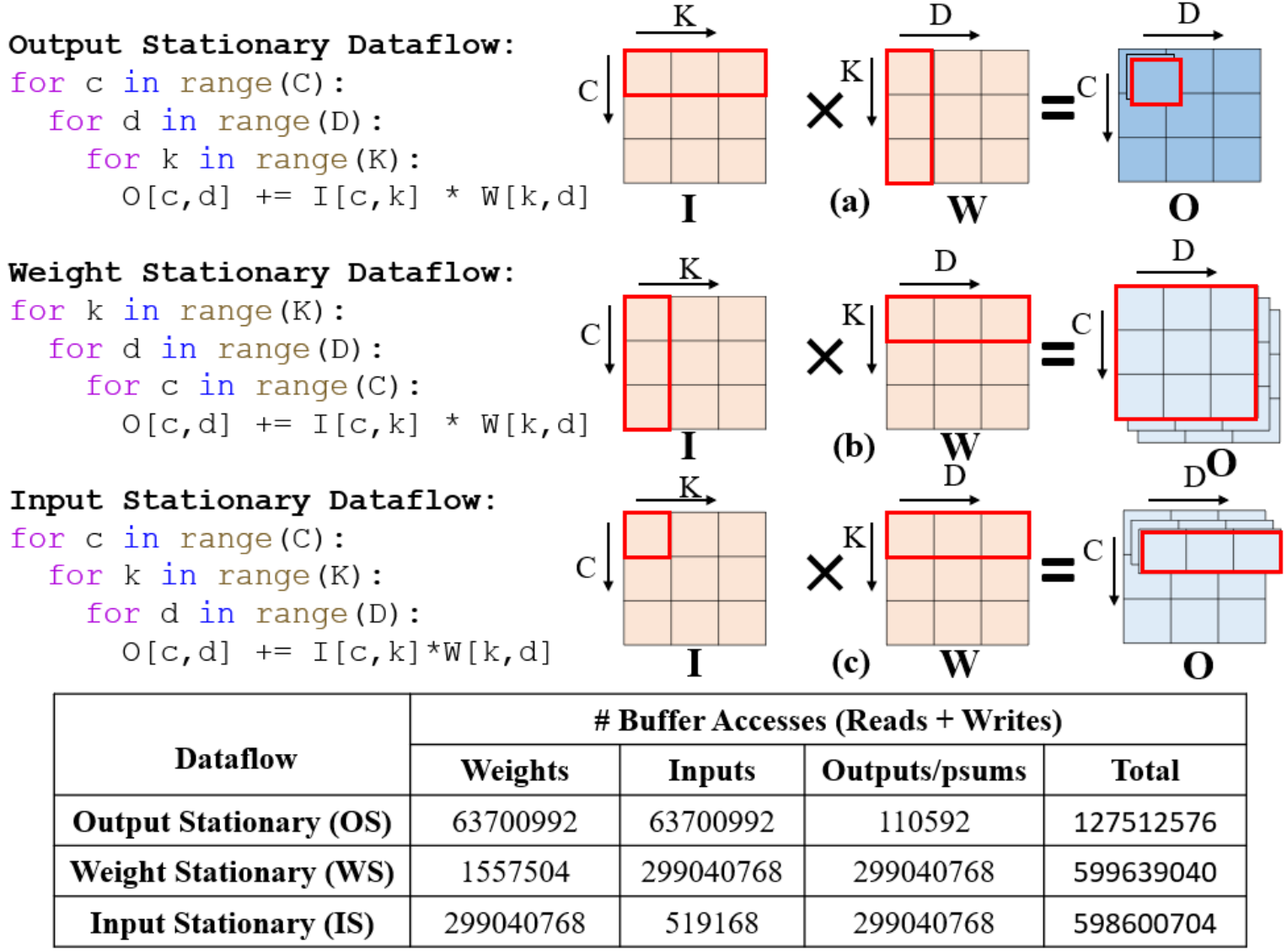}
    \caption{Comparison of CNN dataflow schemes: (a) Output Stationary (b) Input Stationary (c) Weight Stationary. Table reports the buffer accesses required by DPU to process layer 5 of GoogleNet\cite{googlenet}.}
    \label{fig1}
\end{figure} 

\begin{figure}[] 
    \centering
    \includegraphics[scale=0.6]{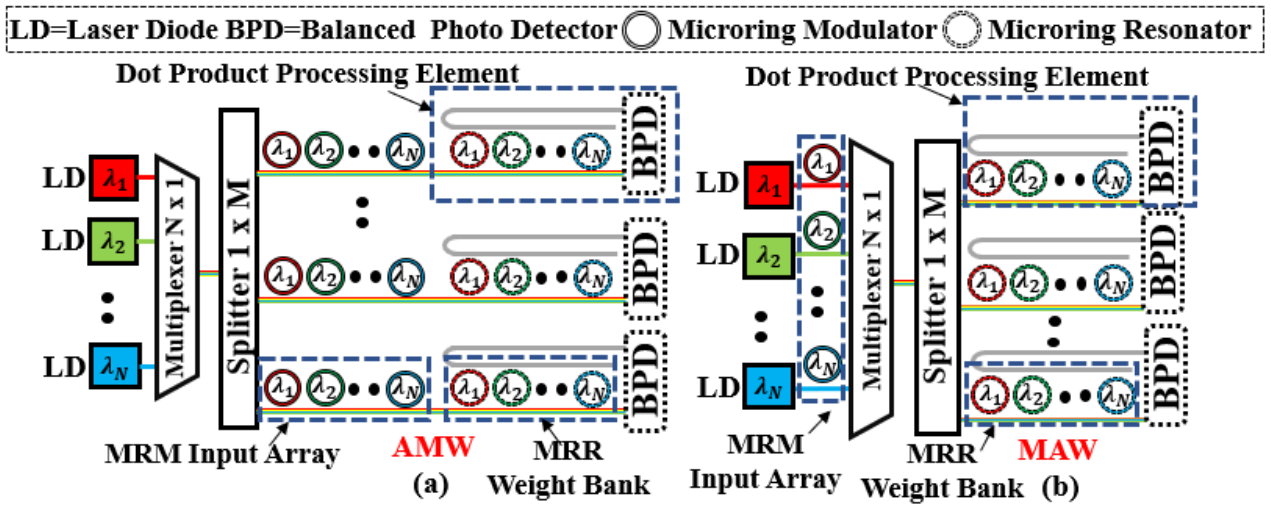}
    \caption{Illustration of common analog optical DPU organizations.(a) AMW DPU (b) MAW DPU.}
    \label{fig2}
\end{figure}

\section{Preliminaries}
\subsection{Processing of CNNs on Hardware}\label{sec21}
In deep CNNs, the major computation requirement arises from convolutional layers. The convolution operations involved in these layers can be converted to General Matrix-Matrix Multiplications (GEMMs) using a relaxed form of the Toeplitz matrix \cite{sze2017efficient}. The Toeplitz matrix \textbf{\textit{I}} of the input tensor of a convolution layer can be obtained with the im2col method (equivalent to the PyTorch's unfold method) \cite{pytorchUnfold}. The weights of the convolution filter tensors are flattened and stacked to form the weight matrix (\textbf{\textit{W}}). The GEMM operation between \textbf{\textit{I}} and \textbf{\textit{W}} gives the resultant output matrix (\textbf{\textit{O}}). On conventional CPUs/GPUs, GEMMs are mapped and executed using basic linear algebra subprograms (BLAS) or Cuda BLAS (cuBLAS) \cite{BLAS,CUBLAS}. For dedicated CNN hardware accelerators, the mapping and execution dataflow of a GEMM function is determined by the loop ordering used for implementing the function. From Fig. 1(a), a GEMM function requires three nested loops involving the dimensions \textit{C}, \textit{K}, and \textit{D} of the matrices \textbf{\textit{I}} and \textbf{\textit{W}}. The order of these dimensions in the nested loops determines the frequency of accesses to the \textbf{\textit{I}}, \textbf{\textit{W},} and \textbf{\textit{O}} matrices. For example, in Fig. 1(b), the given loop order forces the indexes of matrix \textit{W} (i.e., \textit{k} and \textit{d}) to change less frequently compared to indexes of matrix \textbf{\textit{I}} (i.e., \textit{c} and \textit{k}) and 
matrix \textbf{\textit{O}} (i.e., \textit{c} and \textit{d}). Therefore, the data values of the matrix \textbf{\textit{W}} are accessed in a more stationary manner compared to matrices \textbf{\textit{I}} and \textbf{\textit{O}}; hence, this execution is known as weight stationary (\textit{WS}) dataflow. Similarly, the output stationary (\textit{OS}) and input stationary (\textit{IS}) dataflows are illustrated in Fig. 1(a) and Fig. 1(c), respectively.  

In CNN accelerators, the data values of matrices \textbf{\textit{I}}, \textbf{\textit{W}}, and \textbf{\textit{O}} are typically read from and written into a unified buffer \cite{maeri2018}. The number of accesses to this unified buffer plays a significant role in the overall throughput and energy efficiency achieved by the accelerator. The total number of buffer accesses corresponding to a GEMM function changes based on the choice of dataflow. In Fig. \ref{fig1}, the table inset reports the total buffer accesses (reads plus writes) required to finish the execution of the GEMM function corresponding to layer 5 of GoogleNet \cite{googlenet}. The total buffer accesses in the table include accesses to the values of matrices \textbf{\textit{I}}, \textbf{\textit{W}}, and \textbf{\textit{O}}. An output write access is generally performed when the accelerator produces an output value. However, accelerators often produce multiple intermediate results towards a single output value. These intermediate results are known as \textit{psum}s (a shorthand for partial sums), which have to be accumulated together (often using a psum reduction network \cite{maeri2018,sigma2020,tpu}) to calculate the final output value. In case psums are required, the output accesses would involve buffer accesses for both the final output values and \textit{psum}s. As shown in the table inset of Fig \ref{fig1}, the total buffer accesses vary across the \textit{WS}, \textit{IS}, and \textit{OS} dataflows. The \textit{WS} and \textit{IS} dataflows result in the least number of weight and input accesses, respectively, from the unified buffer. Similarly, the \textit{OS} dataflow results in the least output accesses. Atop dataflows, the total access count also depends on the dimensions \textit{C}, \textit{K}, and \textit{D} of the matrices. 

\subsection{Related Work on Optical CNN Accelerators}
Electronic ASICs have traditionally been the preferred choice for implementing CNN accelerators \cite{tpu}. In light of the deceleration of Moore's Law, coupled with the exponential growth in CNN complexity \cite{Baischer2021}, electronic ASIC accelerators are struggling to meet the processing speed and energy efficiency demands for large-scale deployment of complex CNN models. To tackle this challenge, both industry and academic researchers are now investigating innovative more-than-Moore technologies that can offer consistently faster and energy-efficient hardware solutions for CNN acceleration in the foreseeable future. Among various technologies, silicon photonics stands out as a promising candidate, offering unparalleled parallelism, ultra-low latency, and high energy efficiency \cite{cansu2021,albireo,crosslight}. Silicon photonics-based accelerators harness linear photonic phenomena, such as optical transmission and optical signal superposition within photonic integrated circuits \cite{cansu2021,deapcnn}, to accelerate CNN inference. This acceleration results in remarkably fast processing speeds and subnanosecond input-to-output latency, following an O(1) scaling law.

To accelerate CNN inferences with low latency and low energy consumption, prior works proposed various accelerators based on photonic integrated circuits (PICs) (e.g., \cite{holylight,crosslight,deapcnn,cansu2021,sconna}). These accelerators employ PIC-based Dot Product Units (DPUs) to perform multiple parallel dot product operations. Some accelerators implement digital DPUs (e.g., \cite{pixel,albireo,sconna}), whereas some others employ analog DPUs (e.g., \cite{holylight,crosslight,deapcnn, cases2022}). Different DPU implementations employ MRRs (e.g., \cite{holylight,deapcnn,crosslight,tait2020photonic,sconna}) or MZIs (e.g., \cite{cansu2021,mzicomplex2021}) or both (e.g., \cite{pixel}, \cite{albireo}). The analog DPUs can be further classified as incoherent (e.g., \cite{holylight,crosslight,deapcnn, sairamisqed2024}) or coherent (e.g., \cite{coherentrayhamerly2019,coherentZhao2019,coherentnature21}). To set and update the values of the individual input and weights used for vector dot product operations, the incoherent DPUs utilize the analog power amplitudes of optical signals, whereas the coherent DPUs utilize the electrical field amplitude and phase. The coherent DPUs achieve low inference latency, but they suffer from control complexity, high area overhead, low scalability, low flexibility, high encoding noise, and phase error accumulation issues \cite{limitsCohorent}. In contrast, the accelerators based on MRRs-enabled incoherent DPUs achieve better scalability and lower footprint, because they use PICs that are based on compact MRRs \cite{deapcnn}, unlike the coherent DPUs that use PICs based on bulky MZIs. Various state-of-the-art PIC-based optical CNN accelerators are well discussed in survey papers \cite{acceleratorssurvey,acceleratorsurvey2Sudeep,acceleratorsurvey3Bhavin}. Because of the inherent advantages of MRR-enabled incoherent DPUs, there is an impetus to design more energy-efficient and scalable CNN accelerators employing MRR-enabled incoherent (analog) DPUs.

\subsubsection{Organizations of optical incoherent analog DPUs:} S. S. Vatsavai et al. in \cite{cases2022} categorized the organizations of optical analog DPUs from prior works into two groups: Aggregate Modulate Modulate (AMM) and Modulate Aggregate Modulate (MAM) \cite{cases2022}. Here, 'Aggregate' refers to an aggregation of multiple wavelength signals into a single photonic waveguide through wavelength-division-multiplexing (WDM). The first 'Modulate' refers to the modulation of optical wavelength signals with input values, and the second 'Modulate' refers to the modulation (weighting) of input-modulated optical wavelength signals with weight values. To avoid confusion between two 'Modulate' terms, we propose to replace the second 'Modulate' with 'Weight' to imply the weighting of input-modulated signals. Thus, we categorize DPU organizations as Aggregate Modulate Weight (AMW) and Modulate Aggregate Weight (MAW). 

Fig. \ref{fig2} illustrates the schematics of DPU organizations of AMW and MAW categories. In general, these DPUs employ microring modulator (MRM) input arrays and MRR weight banks. In a DPU, inputs are modulated as analog power amplitudes of optical wavelength signals using the MRM input arrays. The individual MRMs of an MRM input array are parallel coupled to a photonic waveguide that carries the multiplexed optical wavelength signals. These optical wavelength signals, after being modulated and multiplexed into the waveguide by the MRM input array, are sent to MRR weight banks. Each MRR in an MRR weight bank is a tunable spectral filter, which consists of an add-drop microring (i.e., a microring coupled to two parallel bus waveguides; one input-through waveguide, and another add-drop waveguide). Each MRR independently controls the transmission of exactly one optical wavelength signal to generate a weighted optical wavelength signal. Each weighted optical wavelength signal is basically a temporal train of optical amplitude symbols carried on an optical wavelength propagating in the waveguide. This amplitude symbol represented the product of the corresponding input and weight values. In a DPU, each waveguide in fact propagates multi-wavelength weighted optical signals, which are multiplexed by the MRM input array, to a balanced photodetector (BPD) implemented at the end of the waveguide. The BPD Balanced photodetection of these weighted, mutually incoherent WDM signals, through the in Figure \ref{fig2}, The BPD produces an electrical current signal, which is a temporal train of electrical current amplitudes. Each current amplitude represents the signed sum of the multi-wavelength temporally coincident optical amplitude symbols that are part of the incident multi-wavelength weighted optical signals \cite{Tait18}. In other words, each current amplitude represents the dot product of the incident wavelength-parallel input and weight vectors, and hence, the electrical current signal at the output of each BPD represents a dot-product signal (i.e., a temporal train of dot-product results). Since a total of \textit{M} BPDs are employed in a DPU (Fig. \ref{fig2}), each DPU generates a total of \textit{M} parallel dot-product signals. Each symbol of a dot-product signal could be either the final output value or a psum. 

\subsection{Motivation}
The CNN accelerators from prior works, which are based on incoherent analog DPUs, have three shortcomings. \textit{First}, the inter-modulation crosstalk in MRM input arrays \cite{cases2022,praneeth4pam} and inter-spectral, electrical, and thermal crosstalk effects in MRR weight banks \cite{Tait18} put forth a strong trade-off between the achievable DPU size (N) (determined by the achieved wavelength parallelism - Fig. \ref{fig2}) and supported bit-precision (B) of AMW and MAW DPUs \cite{lukasscalability,cases2022}. This is because the collective power penalty induced by these crosstalk effects can substantially reduce the available optical power budget in DPUs, which can significantly reduce (i) the tolerance to high optical losses caused by large DPU sizes, and (ii) the dynamic range of optical power required to support large bit-precision. As a result, it is shown that MAW DPUs cannot achieve larger than 44$\times$44 size for $>$4-bit precision \cite{cases2022}.
\textit{Second}, the presence of various crosstalk effects and high spectral sensitivity in the MRR weight banks requires the use of extremely complex control procedures for the actuation of weight values \cite{Tait18, Tait2022}. Such control procedures often employ binary search algorithms\cite{Tait18} or need a feedback control circuit \cite{Tait2022}. This requirement increases the implementation complexity, mandating the weight actuation control to be separate from the required thermal stability control per MRR. Thus, each MRR requires two feedback control circuits, one for weight actuation and one for thermal stabilization. Similarly, each MRM already requires a separate input actuation control due to its high-speed operation (typically >1 gigasybols per second). This would increase the number of required feedback control units per weighted optical signal to four because both the input MRM and weighting MRR would require one feedback control unit each for thermal stabilization and another unit each for value actuation. Each control circuit consumes a significant amount of static power \cite{Tait2022}.  As a result, the generation of each N-sized dot-product at a BPD would increase the total static power consumption by 4N$\times$, diminishing the overall energy efficiency of the DPUs. 
\textit{Third}, the AMW and MAW DPUs from prior works fail to fully leverage the ability of BPDs to accumulate \textit{psums} in situ temporally. This ability endows the BPDs with the potential of eliminating the need to use \textit{psum} buffers and dedicated \textit{psum} reduction networks \cite{maeri2018,sigma2020,albireo}. Failing to leverage this ability significantly increases the latency and energy consumption of the AMW and MAW DPUs from prior works because of their necessity to frequently read and write \textit{psums} into a buffer and to employ dedicated \textit{psum} reduction networks. Our proposed accelerator, HEANA, tackles these shortcomings, as summarized in Section \ref{625}.

\section{Our proposed HEANA Architecture}

\subsection{Overview}
The main processing unit of our HEANA architecture is a dot product unit (DPU), which is illustrated in Fig. \ref{HEANA}. A HEANA DPU consists of a comb laser source \cite{praneeth4pam,Wang23} that emits optical power at a total of \textit{N} distinct wavelengths (i.e., from $\lambda_1$ to $\lambda_N$). The N-wavelength (N-chromatic) optical power sourced from the comb laser is split into a total of \textit{M} waveguide. Each waveguide carries the N-wavelength optical power to a dot product element (DPE).    

\subsection{HEANA's Dot Product Element (DPE)}\label{section3_2}
\begin{figure*}[h!] 
    \centering
    \includegraphics[width = \linewidth]{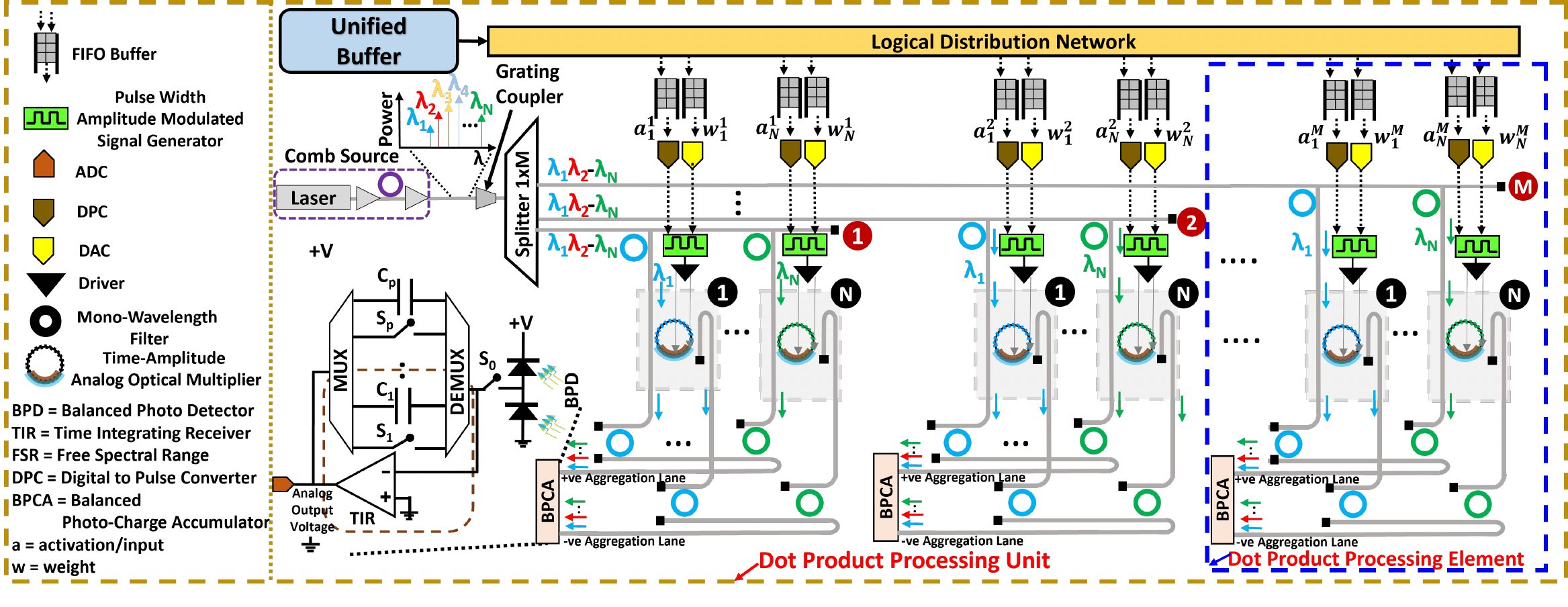}
    \caption{Schematic of the Dot Product Unit (DPU) of our HEANA accelerator.}
    \label{HEANA}
\end{figure*}

From Fig. \ref{HEANA}, a HEANA DPE employs two artifacts: \textit{(i)} a spectrally hitless array of a total of \textit{N} hybrid time-amplitude analog optical modulators (TAOMs), and \textit{(ii)} a balanced photo-charge accumulator (BPCA). A TAOM generates a weighted optical wavelength signal as a temporal sequence of pulse-width-amplitude-modulated (PWAM) symbols. The total optical energy contained in a PWAM symbol generated by a TAOM represents the analog product of one input and one weight value. The BPCA circuit leverages the temporal charge accumulation and incoherent superposition abilities of photodetectors \cite{BhaskaranPCA2022}\cite{alexandermit2022} to generate a signed summation of a large number of temporally and spatially arriving PWAM symbols. This signed summation represents the final dot product result, i.e., a value in the final output matrix. The value \textit{N} here represents the degree of spatial (wavelength) parallelism, which is equal to the number of optical wavelength signals and the number of TAOMs per DPE. The value \textit{N} is also referred to as the size of the DPE. The arrangements of TAOMs and BPCA, along with their structures and operations, are described in the following subsections. 

\subsubsection{Spectrally Hitless Array of N TAOMs:} \label{sec321}
In a DPE, the spectrally hitless array of \textit{N} TAOMs receives N-wavelength optical power from a waveguide fed from the splitter (Fig. \ref{HEANA}). On this waveguide, a total of \textit{N} mono-wavelength filters filter a total of \textit{N} wavelengths individually, and then drop them onto the respective waveguides of \textit{N} TAOMs. Thus, each TAOM operates on a unique optical wavelength signal $\lambda_i$ as discussed in Section \ref{section3_2}.
This arrangement of TAOMs is spectrally hitless \cite{karen2020proceeding} because each TAOM waveguide employs only a single optical wavelength signal. This avoids inter-wavelength interference \cite{modulatorcrosstalk} known as crosstalk from TAOMs operating on adjacent optical wavelength signals. On the contrary, the AMW and MAW DPEs from prior works \cite{holylight,deapcnn,cases2022} typically face such crosstalk because they employ a parallel-coupled array of MRMs and MRRs. This arrangement in an N-sized AMW/MAW DPE causes each of the \textit{N} optical wavelength signals to interact with a total of \textit{N} MRMs as well as \textit{N} MRRs, resulting in a substantial amount of crosstalk noise. To minimize this crosstalk noise, AMW/MAW DPEs typically maintain a wide spectral spacing between adjacent optical wavelength signals. However, this restricts the achievable \textit{N} in the DPEs because a wide wavelength spacing reduces the number of optical wavelength signals that can be spectrally multiplexed within the stringently limited Free Spectral Range (FSR) of the MRMs and MRRs employed in these AMW/MAW DPEs \cite{lukasscalability,cases2022}. In contrast, the spectrally hitless arrangement of TAOMs in HEANA completely eliminates the crosstalk noise at TAOMs, which allows for a narrow wavelength spacing to increase the achievable \textit{N} for HEANA at a given bit precision (more on this in Section \ref{section5}). 

Each spectrally hitless TAOM receives two operands, i.e., input/activation a$_i^M$ and weight w$_i^M$, from corresponding FIFO buffers. The activation and weight values are loaded into the FIFO buffers from the unified buffer via a distribution network based on the selected dataflow. Each TAOM is driven by an electrical pulse width amplitude modulated (PWAM) signal. This signal is electro-optically modulated onto the TAOM's corresponding optical wavelength $\lambda_i$ to generate an optical PWAM signal. Each PWAM symbol of this signal encodes the input a$_i^M$ as its pulse-width and weight w$_i^M$ as its amplitude so that the total optical energy packetized in the PWAM symbol represents the multiplication result between a$_i^M$ and w$_i^M$.
The sign of the multiplication result depends on the sign of w$_i^M$ as activation values in different layers can be considered positive after the application of the ReLU nonlinear function \cite{deapcnn,holylight, crosslight, cases2022, sconna}. This sign is encoded in the form of a balanced PWAM optical symbol at the through and drop ports of the TAOM.
The multiplication results (i.e., balanced PWAM optical symbols) from a total of \textit{N} TAOMs are then dropped into the positive and negative aggregation lanes via a set of mono-wavelength filters \cite{TaitBalancedWDM2012} (Fig. \ref{HEANA}). These aggregation lanes further guide the multiplication results to the BPCA for accumulation. The BPCA, in each symbol cycle, receives \textit{N} individual multiplication results concurrently (i.e., \textit{N} PWAM symbols), incoming from \textit{N} TAOMs on \textit{N} parallel wavelengths. The BPCA transduces the total optical energy packetized in the \textit{N} multiplication results (PWAM symbols) arriving at the BPCA into an analog voltage amplitude, which represents a \textit{psum} of the final dot product result (a value in the output matrix \textbf{\textit{O}} in Fig. \ref{fig1}). Depending on the selected dataflow, a specific capacitor in our BPCA circuit is utilized to transduce and temporarily store the \textit{psum}. The use of capacitors in the BPCA also allows bufferless temporal accumulation of subsequently arriving \textit{psums}. The design and operation of TAOM and BPCA are explained next.   

\begin{figure*}[h!]
    \centering
    \includegraphics[width = \linewidth]{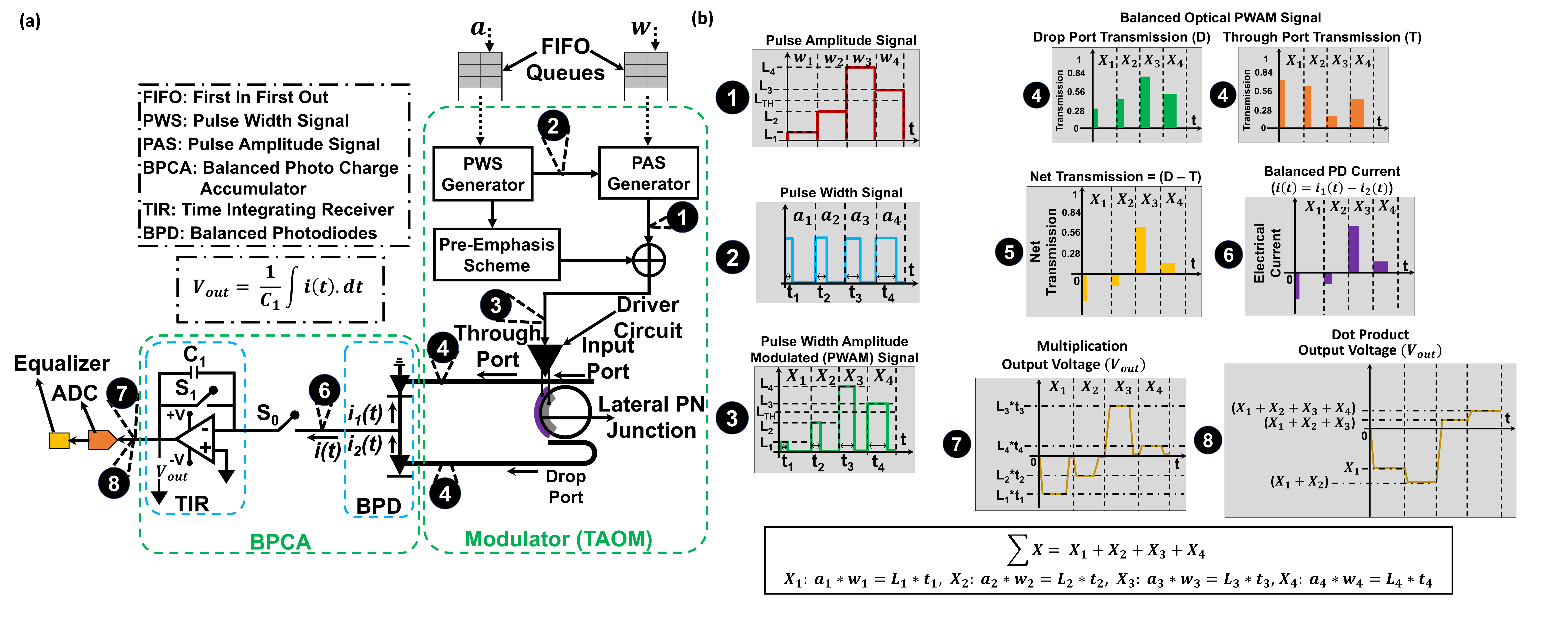}
    \caption{(a) Structure of our microring modulator (MRM) based hybrid time-amplitude analog optical modulator (TAOM) connected to a balanced photocharge accumulator (BPCA) and (b) representation of analog signals (optical and electrical) at different stages of TAOM.}
    \label{TAOM}
\end{figure*}

\subsubsection{Design of Hybrid Time-Amplitude Analog Optical Modulator (TAOM):}
Fig. \ref{TAOM}(a) depicts the schematic of our invented TAOM when it is connected to a balanced photo-charge accumulator (BPCA) unit. As illustrated, our TAOM is basically an add-drop Microring Modulator (MRM) with an embedded lateral PN junction that operates in the forward bias condition. The MRM's peripheral circuitry consists of two queues of FIFO buffers, in which one of them stores the input values (from the input matrix shown in Fig. \ref{fig1}) and the other one stores the weight values (from the weight matrix shown in Fig. \ref{fig1}), both in the digital binary-radix number format. The FIFO queue for inputs connects to a pulse width signal (PWS) generator and the FIFO queue for weights connects to a pulse amplitude signal (PAS) generator. The output of the PWS is split into two parts: one part is directed to the pre-emphasis scheme, whereas the other part is provided as a reference to the PAS generator. Subsequently, the output of the PAS generator and the output of the pre-emphasis scheme are combined through a current-mode mixer, and the resulting output is a pulse-width-amplitude-modulated (PWAS) signal. Our TAOM requires simpler synchronization of PWS and PAS signals. The synchronization occurs locally per TAOM with clock sharing, minimizing signal jitter and skew and making our design easier to implement. For a complete understanding of the generation of PWAM signals and the underlying circuitry, we direct the readers to \cite{kim2022high,yang2008pwm}. This PWAS signal is routed to a driver circuit. The output of the driver circuit is provided as an electrical bias to the PN junction of the MRM. The output of the MRM (TAOM) is connected to a balanced photo charge accumulator (BPCA) circuit. 

\subsubsection{Balanced Photocharge Accumulator (BPCA)}
Our BPCA circuit is collectively inspired by the time integrating receiver (TIR) design from \cite{alexandermit2022,tirdatasheet} and the photodetector-based optical pulse accumulator design from \cite{BhaskaranPCA2022}. As illustrated in Fig. \ref{TAOM}(a), a BPCA circuit employs two photodiodes, each connected to the drop and through ports of the MRM. These photodiodes are interlinked in a balanced configuration, commonly referred to as a balanced photodiode (BPD) configuration. The BPD is connected to a TIR via a switch (S$_{0}$). The TIR comprises an amplifier and a feedback capacitor/switch (S$_{1}$) pair (Fig. \ref{TAOM}). It functions as a current-to-voltage converter circuit by integrating the incoming electrical current over a period. This ensemble of the BPD and TIR makes the BPCA capable of performing temporal and spatial accumulations (this will be explained in upcoming subsections). Subsequently, the output of the TIR is connected to an analog-to-digital converter (ADC) and an equalizer.

\subsubsection{Operation of Integrated TAOM-BPCA Circuit}
Figure \ref{TAOM}(b) illustrates the sequential processing of electrical and optical signals at various stages within our integrated TAOM-BPCA circuit, demonstrating the effective execution of the multiplications and temporal accumulations. As illustrated, the FIFO queue for weight values feeds into the PAS generator, which converts the incoming sequence of digital weight values into a sequence of analog pulse amplitude symbols. This sequence is also called a pulse amplitude signal ((see \filledcircled{1} in Figs. \ref{TAOM}(a) and \ref{TAOM}(b))). Similarly, the FIFO queue for input values feeds the PWS generator, which converts the incoming sequence of digital input values into a sequence of analog pulse-width-modulated symbols. This sequence is called a pulse width signal (see \filledcircled{2} in Figs. \ref{TAOM}(a) and \ref{TAOM}(b)). The PWS output is divided, with one part directed to the pre-emphasis scheme, while the other part serves as a reference for the PAS generator. The output of the PAS generator (current-mode DAC), when mixed with the output of the pre-emphasis scheme, produces a sequence of pulse width amplitude modulated (PWAM) symbols ((see \filledcircled{3} in Fig. \ref{TAOM}(b))). This sequence is called PWAM signal. For a complete understanding of the generation of PWAM signals and the underlying circuitry, we direct the readers to \cite{kim2022high,yang2008pwm}. The PWAM signal is fed to a driver circuit, as shown in Fig. \ref{TAOM}(a). From the driver circuit, the PWAM signal is provided as an electrical input to the PN junction of the MRM. 

This input PWAM signal induces free-carrier plasma dispersion in the MRM, which enables the MRM to dynamically adjust the transmission characteristics of the incoming wavelength channel from an external laser source. This action converts the input electrical PWAM signal into a balanced optical PWAM signal. Here, a balanced optical PWAM signal implies that the original electrical PWAM signal is encoded into optical transmissions simultaneously at both the drop and through ports of the MRM ((See \filledcircled{4} in Fig. \ref{TAOM}(b))). For each symbol of this balanced optical PWAM signal, the amount of transmission at the through and drop ports of the MRM depends on the amplitude level of each symbol relative to the threshold level in the electrical PWAM signal (L$_{TH}$ in \filledcircled{3} of Fig. \ref{TAOM}(b)). For instance, the amplitudes of symbols \textbf{X$_{1}$} and \textbf{X$_{2}$} in \filledcircled{3} of Fig. \ref{TAOM}(b) are below the defined threshold level (L$_{TH}$). Therefore, for symbols \textbf{X$_{1}$} and \textbf{X$_{2}$} in \filledcircled{4} of Fig. \ref{TAOM}(b), the transmission at the drop port of the MRM is lower than the transmission at the through port. As a result, the net transmission, represented as the difference in transmission between the drop and through ports of the MRM (\textbf{X$_{1}$} and \textbf{X$_{2}$} in \filledcircled{5} of Fig. \ref{TAOM}(b)), is negative for these symbols. On the other hand, the amplitudes of symbols \textbf{X$_{3}$} and \textbf{X$_{4}$} in \filledcircled{3} of Fig. \ref{TAOM}(b) are above the defined threshold level (L$_{TH}$). Therefore, for the symbols \textbf{X$_{3}$} and \textbf{X$_{4}$} in \filledcircled{4} of Fig. \ref{TAOM}(b), the transmission at the drop port of the MRM is higher than the transmission at the through port. As a result, the net transmission (\textbf{X$_{3}$} and \textbf{X$_{4}$} in \filledcircled{5} of Fig. \ref{TAOM}(b)), is positive for these symbols. Each such symbol of a balanced optical PWAM signal (such as X$_1$ in \filledcircled{4} of Fig. \ref{TAOM}(b)) packetizes certain optical energy that is proportional to the analog product of the corresponding input (\textit{a}) and weight values (\textit{w}). For example, the energy packetized in symbol \textbf{{X$_{1}$}} represents \textbf{{a$_{1}$}}*\textbf{{w$_{1}$}} (or) \textbf{{L$_{1}$}}*\textbf{{t$_{1}$}} in Fig. \ref{TAOM}(b).  This balanced optical PWAM signal at the output of the TAOM is fed into the BPCA circuit. 

Within the BPCA circuit, the BPD transduces the incoming optical pulse sequence (\textbf{X$_{1}$},...,\textbf{X$_{4}$} in \filledcircled{5} of Fig. \ref{TAOM}(b)) from the MRM into a series of differential electrical current amplitudes (\filledcircled{6} in Fig.\ref{TAOM}(b)). Here, the differential electrical current amplitude corresponding to each symbol is proportional to the net transmission of the respective optical PWAM symbol (\filledcircled{5} in Fig. \ref{TAOM}(b)). Moreover, the multiplication magnitude corresponding to each symbol is encoded as the area under the curve of the differential electrical current symbol. The direction of the electrical current symbol (incoming to the BPD and outgoing from the BPD) represents the sign of the multiplication. This series of differential electrical current symbols is directed towards the TIR via the switch S$_{0}$. The integration of TIR with the BPD introduces a distinctive versatility that enables us to operate our integrated TAOM-BPCA circuit in one of the two distinct modes: (i) multiplier mode or (ii) multiplier and temporal accumulator mode. These modes are explained next. 

\textbf{(i) Multiplier Mode:} For this mode, the TIR's sampling speed is matched to the arrival rate of the incoming differential electrical current symbols. At the beginning of each symbol period, opening the switch S$_{1}$ allows the electrical current symbol corresponding to that period to linearly charge the feedback capacitor C$_{1}$ of the TIR circuit. This linear charging continues for a duration equal to the pulse width of the electrical current symbol. Consequently, the accumulated charge, and therefore, the analog voltage accrued across the feedback capacitor C$_{1}$ of the TIR for that symbol period represents the multiplication result related to the corresponding optical PWAM symbol (see \filledcircled{7} in Fig. \ref{TAOM}(b)). Notably, the polarity of the incoming differential electrical current pulse indicates the sign of the multiplication result; therefore, the polarity of the accrued analog voltage across the TIR's feedback capacitor becomes negative if the incoming electrical current has negative polarity. Once the accrued analog voltage is stable, it is sampled and sent to an analog-to-digital converter (ADC). Then, at the end of the symbol period, closing the switch S$_{1}$ allows resetting the charge and voltage on the feedback capacitor to be zero, to prepare the TIR for the next symbol period. 

For example, consider the symbol \textbf{X$_{1}$} in \filledcircled{7} of Fig. \ref{TAOM}(b). The feedback capacitor of the TIR circuit linearly charges until it reaches an analog voltage level of (L$_{1}$*t$_{1}$), which represents the signed multiplication result of the symbol \textbf{X$_{1}$}. After the analog voltage level on the capacitor reaches (L$_{1}$*t$_{1}$) for the symbol \textbf{X$_{1}$} and the capacitor reaches a steady state, the accrued analog voltage is sampled and then sent to the ADC and equalizer for further processing, before the capacitor is made to discharge (reset) by closing the switch S$_{1}$. Here, since the polarity of the differential electrical current corresponding to symbol \textbf{X$_{1}$} is negative, the accrued analog voltage on the TIR's feedback capacitor is also negative, as shown for symbol \textbf{X$_{1}$} in \filledcircled{7} of Fig. \ref{TAOM}(b). On the other hand, if the polarity of the incoming differential electrical current pulse is positive, the accrued analog voltage on the TIR's feedback capacitor is also positive, which is illustrated for symbol \textbf{X$_{3}$} in \filledcircled{7} of Fig. \ref{TAOM}(b). Thus, in this mode, the TAOM-BPCA circuit acts as a multiplier that can produce multiplication results at a fast speed. 

\textbf{(ii) Multiplier and Temporal Accumulator mode:}
For this case, the TIR's sampling speed is set to be very low compared to the arrival rate of the incoming differential electrical current symbols. Therefore, the series of differential electrical current symbols arriving at the TIR can sequentially charge the TIR's capacitor so that the net accumulated charge and, consequently, the analog voltage accrued on the capacitor over multiple symbol periods provides the signed sum of the individual multiplication results corresponding to different symbols. Thus, this operation essentially performs a temporal accumulation of multiplication results (products). This operation is depicted in \filledcircled{8} of Fig. \ref{TAOM}(b), where the charge accumulates over time based on the incoming electrical current symbols(\textbf{X$_{1}$},...,\textbf{X$_{4}$} in \filledcircled{6} of Fig. \ref{TAOM}(b)), and consequently, the resulting analog voltage accrued on the TIR's capacitor signifies the temporal accumulation result (i.e., \textbf{X$_{1}$}+...+\textbf{X$_{4}$}). 
Moreover, if the incoming differential electrical current pulses to the TIR circuit include both positive and negative polarities, the resultant analog voltage accumulated on the capacitor over time, representing the temporal accumulation operation, is a summation of positive and negative voltages. 
The temporal accumulation operation illustrated in \filledcircled{8} of Fig. \ref{TAOM}(b) is a summation of both negative and positive voltages. The first two symbols of the incoming current signal (i.e., \textbf{X$_{1}$} and \textbf{X$_{2}$} in \filledcircled{6} of Fig. \ref{TAOM}(b)) have negative polarity. Therefore, in \filledcircled{8} of Fig. \ref{TAOM}(b), the net accrued voltage on the capacitor at the end of the second symbol period has negative polarity. The magnitude of this voltage represents \textbf{X$_{1}$}+\textbf{X$_{2}$}. This is because, unlike multiplier mode, the switch S$_{1}$ is not closed every symbol period (rather it is kept open) during the operation of this multiplier+temporal accumulator mode. As a result, the accrued voltage does not return to zero at the end of every period; rather, it builds on top of the voltage level accrued in the previous period. In \filledcircled{6} of Fig. \ref{TAOM}(b), the collective magnitude of the differential electrical current corresponding to the symbols \textbf{X$_{3}$} and \textbf{X$_{4}$} is high compared to that of the symbols \textbf{X$_{1}$} and \textbf{X$_{2}$}. Therefore, the resultant voltage accrued on the capacitor has positive polarity after all four symbol periods have elapsed. This voltage corresponds to the result of the temporal accumulation of incoming PWAM symbols. Since each PWAM symbol encodes a multiplication result, this temporal accumulation result is basically a temporal dot-product result. This dot-product result can be collected by sampling the accrued voltage and then directing it to the ADC and equalizer for further processing. After the desired number of periods for the temporal accumulations have elapsed, the switch S$_{1}$ is closed to reset/discharge the capacitor, to prepare the circuit for the next temporal accumulation. 

\begin{figure*}[h!]
    \centering
    \includegraphics[scale=0.2]{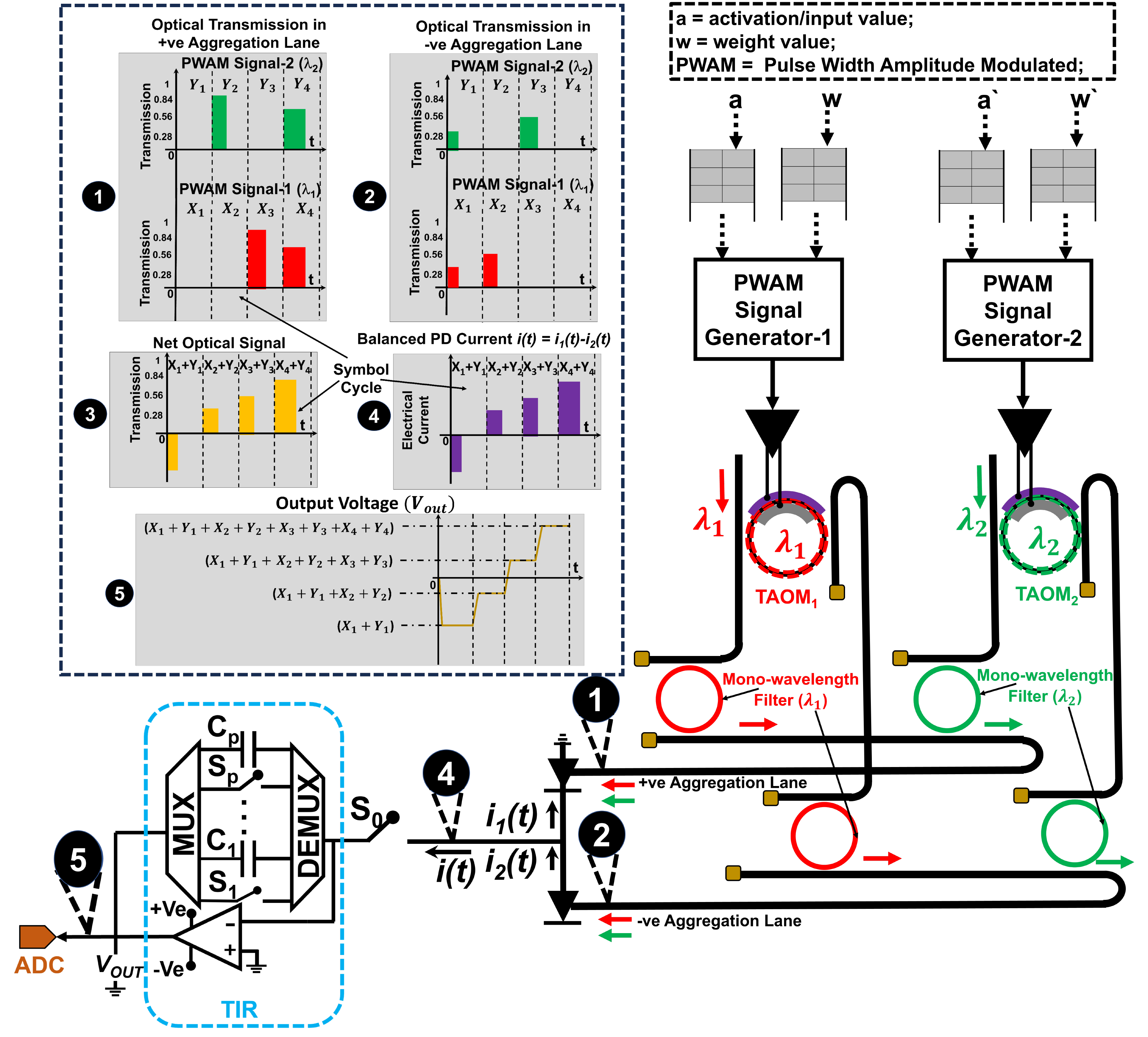}
    \caption{HEANA DPE consisting of two spectrally hitless TAOMs, connected to our BPCA circuit. The inset showcases analog representations of signals (both optical and electrical) at various stages of our DPE}
    \label{DPE}
\end{figure*}

\subsubsection{Spatio-Temporal Multiply-Accumulate Operations in HEANA DPE:}\label{325} Previously, we demonstrated that the ensemble of one TAOM and one BPCA can be used to perform temporal multiply-accumulate operations. In this subsection, we extend that idea and demonstrate that an ensemble of multiple TAOMs and a shared BPCA, which creates a HEANA DPE by employing wavelength-division-multiplexing, can be used to perform spatio-temporal multiply-accumulate operations. To understand this, consider Fig. \ref{DPE} that illustrates the functioning of an example HEANA DPE comprised of an ensemble of two spectrally hitless TAOMs (TAOM$_1$ and TAOM$_2$) and a shared BPCA. From what we know about TAOM-BPCA ensemble discussed earlier in Section 3.2.4 with respect to Fig. \ref{TAOM}, TAOM$_1$ and TAOM$_2$ in the HEANA DPE of Fig. \ref{DPE} generate balanced optical PWAM signals that are carried onto the dedicated optical wavelengths $\lambda_1$ and $\lambda_2$ respectively. Both of these balanced optical  PWAM signals (shown as optical transmissions in \filledcircled{1} and \filledcircled{2} of Fig. \ref{DPE}) are multiplexed into the positive and negative accumulation lanes (lanes are waveguides) using mono-wavelength filters. The aggregation lanes guide these balanced optical PWAM signals to the shared BPCA. During each symbol cycle, the BPD of the BPCA performs a balanced incoherent superposition (signed summation) of all the balanced optical PWAM symbols that arrive during the symbol cycle. Consequently, the balanced incoherent superposition first enables the creation of a net optical signal (see \filledcircled{3} in Fig. \ref{DPE}), and then, allows this net optical signal to be transduced to generate a balanced photocurrent signal (see \filledcircled{4} in Fig. \ref{DPE}). The area under the curve of every balanced photocurrent symbol in \filledcircled{4} of Fig. \ref{DPE} gives a spatial accumulation result (a spatial sum) of the PWAM symbols incident during the corresponding symbol cycle. The polarity of each balanced photocurrent symbol gives the sign of the corresponding spatial sum. Thus, the BPD of the BPCA enables spatial accumulation of incident PWAM symbols. Since all PWAM symbols are multiplication results, their spatial accumulation at the BPCA can also be referred to as spatial multiply-accumulate (MAC) operation or spatial dot-product operation. 

This balanced photocurrent signal (which can also be called photocurrent-based spatial MAC signal) produced by the BPD of the BPCA is sent to the TIR of the BPCA, where it can be further processed in two different ways. \textit{First}, if the sampling rate of the TIR is kept equal to the symbol rate of the incoming balanced photocurrent signal, the TIR simply converts the photocurrent-based spatial MAC signal into a voltage-based spatial MAC signal. In this signal, each symbol simply is a voltage level representing a spatial dot-product result. \textit{Second}, if the sampling rate of the TIR is kept to be an integer multiple of the symbol rate of the incoming photocurrent signal, the TIR enables the gradual integration (temporal accumulation) of the individual symbols of the photocurrent-based spatial MAC signal to provide a single voltage level as the final output that represents the temporal sum of all the individual symbols (spatial dot-product results) of the spatial MAC signal. This occurs due to the same operational characteristics of the TIR as discussed in Section 3.2.4-(ii). Thus, the BPCA (BPD + slowly sampled TIR) enables spatio-temporal MAC or dot-product (i.e., temporal accumulations of spatial dot-product results) in the HEANA DPE.

This spatio-temporal accumulation capability of HEANA DPE is clearly illustrated in \filledcircled{5} of Fig. \ref{DPE}. During the first symbol cycle, X$_1$ and Y$_1$ are spatially accumulated resulting in an accrued \textbf{V$_{out}$} that is proportional to (\textbf{X$_1$} + \textbf{Y$_1$}). Here, the cumulative polarity of (\textbf{X$_1$} + \textbf{Y$_1$}) is negative, thereby resulting in negative \textbf{V$_{out}$}. In the next symbol cycle, the balanced photocurrent generated during the cycle corresponds to spatially accumulated symbols (\textbf{X$_2$} + \textbf{Y$_2$}). This balanced photocurrent accrues voltage on top of the  \textbf{V$_{out}$} accrued in the previous cycle, resulting in updated V$_{out}$ that is proportional to (\textbf{X$_1$} + \textbf{Y$_1$} + \textbf{X$_2$} + \textbf{Y$_2$}). Thus, the temporal accrual of V$_{out}$ over all four symbol cycles enables the final result to be (\textbf{X$_1$} + \textbf{Y$_1$ }+ \textbf{X$_2$} + \textbf{Y$_2$} + \textbf{X$_3$} + \textbf{Y$_3$} + \textbf{X$_4$} + \textbf{Y$_4$}). The polarity of V$_{out}$ depends on the net polarity of the final result. The final V$_{out}$ from the BPCA circuit is given as input to an ADC to produce the final output in the digital format. 

A HEANA DPE can be an ensemble of a shared BPCA and a large number of TAOMs; the number of TAOMs per HEANA DPE is not limited to two. Therefore, this spatio-temporal accumulation capability of HEANA DPE can essentially enable the processing of very large-sized dot-products both spatially as well as temporally. In addition, as shown in Fig. \ref{DPE}, the TIR in the BPCA of a HEANA DPE contains a total of p independently operable capacitors. Each HEANA DPE, while leveraging its spatio-temporal accumulation capability, can intermittently switch among these p capacitors by orchestrating the closing and opening of respective switches to facilitate efficient execution of various dataflows. A detailed explanation of this functionality is provided in Section \ref{section4}.

Furthermore, We model our TAOM unit using photonics foundry-validated simulation tools from ANSYS/Lumerical \cite{lumerical_2021}. Here, we perform a time-domain (transient) analysis to evaluate the performance of our TAOM in terms of accuracy and precision for different values of optical power and sample rates. A detailed discussion of this analysis is provided in the next subsection.  

\begin{figure} 
    \centering
    \includegraphics[width = \linewidth]{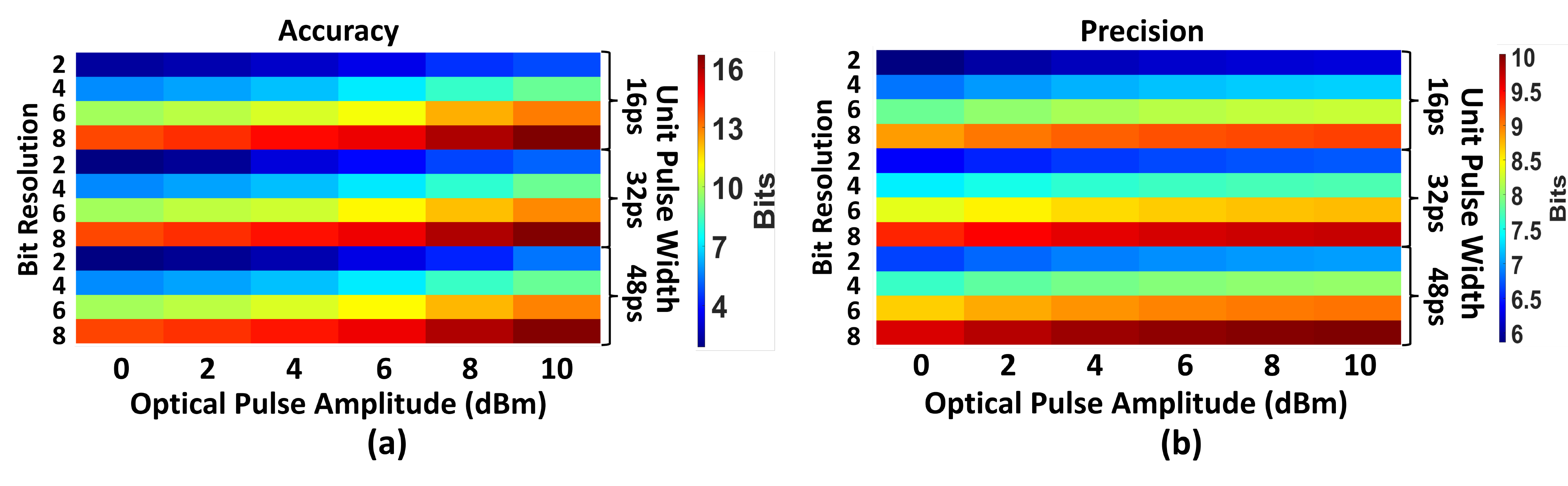}
    \caption{Colormap plots that depict the (a) accuracy and (b) precision of our TAOM for different values of input optical power, sample rate and step size/time interval between the time-analog signals}
    \label{fig5}
\end{figure} 

\subsubsection{Performance Analysis of TAOM:}\label{sec323}
We evaluate the performance of our TAOM in terms of accuracy and precision. To measure the accuracy of our TAOM, we calculated the logarithmic transformation of the inverse of the normalized mean absolute error (MAE) ((log${_2}$(1/(MAE)))) between the actual voltages (i.e., the voltages across the capacitor (C${_1}$) in the BPCA for different multiplication operations that are extracted from simulations) and the commanded/targeted voltages for different multiplication operations.. We represent accuracy in terms of bits by considering various input optical pulse amplitudes and bit resolution values. we considered four values of bit resolution  (2, 4, 6, 8-bits) and three unit sizes of pulse widths (16ps, 32ps, 48ps). If the unit size of pulse width is 16 ps, the input values of unity would be represented as a 16 ps wide pulse. The colormap plot in Fig. \ref{fig5}(a) illustrates the accuracy evaluated for different combinations of optical pulse amplitudes and pulse widths. Additionally, we evaluated precision for these combinations using equations provided in \cite{lukasscalability}. The corresponding colormap plot for precision can be found in Fig. \ref{fig5}(b).

As depicted in the accuracy and precision colormap plots (Fig. \ref{fig5}), for a given bit resolution and pulse width for a unity input value, the accuracy and precision of our TAOM increase when the input optical pulse amplitude increases from 0 dBm to 10 dBm. This is because the optical pulse amplitude is basically representative of the optical power signal, and the increase in the input optical pulse amplitude means an increase in the optical power signal. This improves the signal-to-noise ratio, leading to better accuracy and precision for our TAOM. Similarly, for a given input optical pulse amplitude, the precision of our TAOM increases when the pulse width of a unity input value increases. Furthermore, for a given input optical pulse amplitude and a unit pulse width, the accuracy and precision of our TAOM increase with the increase in bit resolution. These results imply that it is possible to achieve accuracy of as high as 16-bit and precision of as high as 10-bit with our TAOM. These accuracy and precision values for our TAOM are highly competitive compared to the values achievable by analog-only photonic incoherent multipliers (or weight banks) from prior works \cite{Tait2022}\cite{Zhang22Optica}.

\begin{figure}[h]
  \centering
  \includegraphics[width=\linewidth]{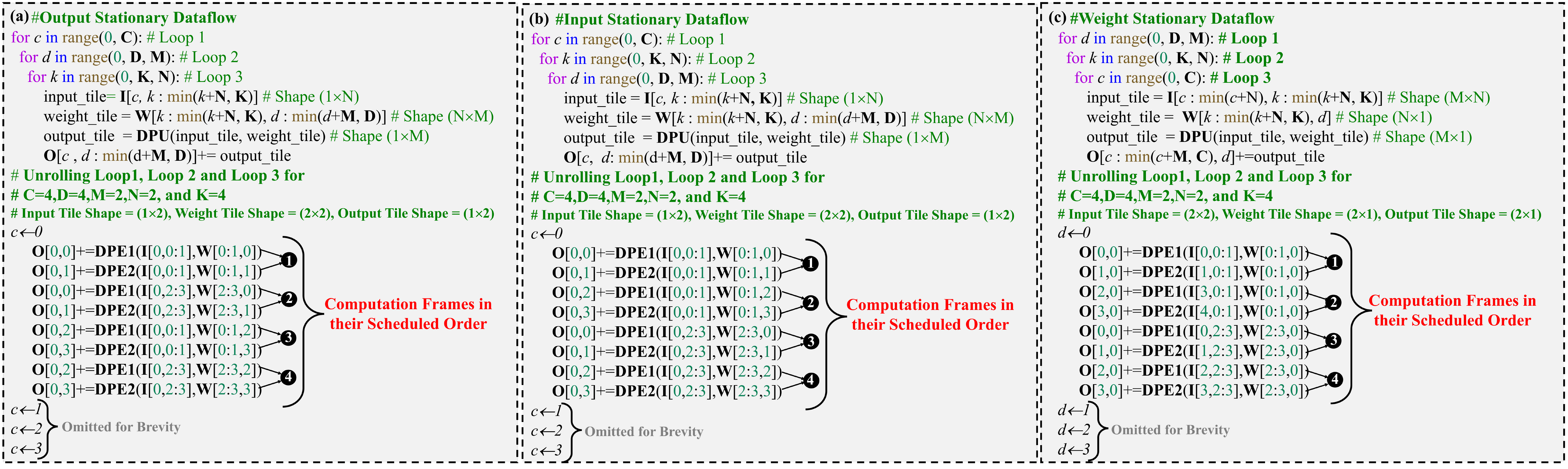}
  \caption{ Loops involved in GEMM operation between input \textbf{\textit{I}} and weight \textbf{\textit{W}}, mapped onto a photonic DPU, with (a) Output Stationary Dataflow (b) Input Stationary Dataflow (c) Weight Stationary Dataflow. DPU consists of 2 DPEs (M=2), and each DPE can perform dot product operation of size N=2. For output stationary dataflow and input stationary dataflow the loops are unrolling to show the computations involved in the evaluation of output \textit{\textbf{O}} row 1  whereas for weight stationary dataflow evaluation of output \textit{\textbf{O}} column 1 is shown. Note that each DPU call performs two dot product operations simultaneously employing the two DPEs. Scheduling order of computation frames for the last three iterations of outermost loops are omitted for brevity}  
  \label{loopunrolling}
  \end{figure}

\section{Mapping of Different Dataflows on HEANA DPU}\label{section4}
The mapping and execution dataflow of a GEMM operation is determined by the loop order, as discussed in Section \ref{sec21}. A GEMM operation requires three nested loops involving the dimensions \textit{C}, \textit{K}, \textit{D} of the matrices \textbf{\textit{I}} and \textit{\textbf{W}} (Section \ref{sec21}). The GEMM operation between matrices \textbf{\textit{I}} and \textit{\textbf{W}} basically requires a total of \textit{C}$\times$\textit{D} \textit{K}-sized dot products. Fig. \ref{loopunrolling} illustrates the loops used for mapping the GEMM operation between \textit{\textbf{I}} and \textbf{\textit{W}} onto a HEANA DPU consisting of \textit{M} DPEs, where each DPE can perform a dot product operation of size \textit{N}. Here, \textit{M} also determines the number of parallel dot product operations that can be implemented on the DPU. The supported size \textit{N} of the DPEs is often less than the required dot product size \textit{K} for implementing GEMM-converted convolutions \cite{cases2022}. In the wake of such size mismatch between the hardware and mapped matrices, to make the required dot products amenable to hardware implementation, the input and weight matrices (\textit{\textbf{I}} and \textbf{\textit{W}}) are mapped to the hardware in smaller blocks by tiling them. The \textit{M} and \textit{N} parameters of the DPU along with the execution dataflow (loop order) determine the shape of the input and weight tiles. These tiles are mapped both temporally and spatially onto the DPEs to compute blocks of output values as output tiles and facilitate the completion of the GEMM operation. In Fig. \ref{loopunrolling}, the innermost loop of each dataflow illustrates how the tile shapes are determined for the input, weight, and output tiles. For the output stationary dataflow illustrated in Fig. \ref{loopunrolling}(a), the input tile shape is \textit{1$\times$N}, whereas for the weight stationary dataflow shown in Fig. \ref{loopunrolling}(c), the input tile shape is \textit{M$\times$N}. Similarly, the shape of the weight and output tiles vary between the output stationary and weight stationary dataflows. Thus, the choice of execution dataflow has an impact on the tiling of \textit{\textbf{I}}, \textit{\textbf{W}}, and \textit{\textbf{O}} matrices of the GEMM operation.    

The choice of execution dataflow (i.e., loop order) for a given \textit{M} and \textit{N} parameters also determines the computation order for output tiles. In Fig. \ref{loopunrolling}, for each dataflow, the loops are unrolled for DPU with parameters \textit{M=2} and \textit{N=2} for input matrix \textbf{\textit{I}} (with dimensions \textit{C}$\times$\textit{K}=4$\times$4) and weight matrix \textbf{\textit{W}} (with dimensions \textit{K}$\times$\textit{D}=4$\times$4). After unrolling, each iteration of the outermost loop involves four computation frames (\filledcircled{1},\filledcircled{2},\filledcircled{3}, and \filledcircled{4}). A computation frame is comprised of dot product operations that can be mapped onto a single DPU in a parallel manner.  The number of computation frames involved in each iteration of an outermost loop can be derived using \#$ComputationFrames$=(ceil($D/M$) $\times$ ceil($K/N$)). For each computation frame, the input and weight tiles are mapped onto a single DPU to perform parallel dot product operations toward computing a single output tile. The DPU employs two DPEs (DPE1 and DPE2) to perform two dot product operations in parallel. In Fig. \ref{loopunrolling}(a), for computation frame \filledcircled{1}, the output stationary dataflow computes partial output values of \textit{\textbf{O}}[0,0] at DPE1 and \textbf{\textit{O}}[0,1] at DPE2. These output values (\textit{\textbf{O}}[0,0] and \textbf{\textit{O}}[0,1]) belong to row 1 of output matrix \textit{O}. In contrast, computation frame \filledcircled{1} of weight stationary dataflow shown in Fig. \ref{loopunrolling}(c) computes output values \textit{\textbf{O}}[0,0] at DPE1 and \textbf{\textit{O}}[1,0] at DPE2, which belong to the column 1 of the matrix \textit{O}. Furthermore, among the output stationary and input stationary dataflows, an identical computation frame can involve the computation of different output values belonging to the same output row of the matrix\textit{\textbf{O}}. For instance, in Fig. \ref{loopunrolling}(a) and Fig. \ref{loopunrolling}(b),  computation frame \filledcircled{2} for both the output stationary and input stationary dataflows compute the output values corresponding to the row 1 of matrix \textbf{\textit{O}}; however, the output stationary dataflow computes the output values \textit{\textbf{O}}[0,0] and \textbf{\textit{O}}[0,1], whereas the input stationary dataflow computes different output values \textit{\textbf{O}}[0,2] and \textbf{\textit{O}}[0,3] belonging to the same row. Thus, the computation order of output tiles varies with the choice of execution dataflow. This emphasizes the critical role of the dataflow in shaping the execution order of the GEMM operation. 

In the next three subsections, we will present a detailed explanation and a comprehensive comparison of GEMM operations' mapping onto HEANA and AMW \cite{deapcnn} architectures. We will consider output stationary (\textit{OS}), input stationary (\textit{IS}), and weight stationary (\textit{WS}) dataflows. \textbf{Important Definitions:} To help with our explanation, we introduce two tiling directives: temporal switching (\textit{ts}) and temporal folding (\textit{tf}) to elucidate the temporal mapping of input and weight tiles. Both the temporal switching and temporal folding of input/weight tiles mean that the input/weight tiles that are mapped onto the DPU change over time in a pipelined manner. The unit amount of time for which the input/weight tiles are held in the DPU for processing before folding (switching) in a pipelined manner is referred to as a folding (switching) cycle. Temporal switching and temporal folding, however, differ in the way that they impact the mapping and processing of the output tile. Temporal switching of the input/weight tile alters the output tile being computed at each switching cycle. In contrast, employing temporal folding means that the computation of the same output tile is folded across multiple cycles, with each folding period generating partial results for each value in the output tile. The lengths of the switching and folding cycles may differ across different dataflows as well as between input and output tiles. 

\textbf{Example DPU and GEMM configurations for explanation of dataflow mappings:} For our explanation of various dataflow mappings in the following subsections, we assume the availability of a single DPU composed of 2 DPEs (\textit{M}=2), where each DPE has a size of \textit{N}=2. Each DPE is equipped with one BPCA containing 2 capacitors (\textit{p}=2) to facilitate in-situ spatio-temporal accumulations (as discussed earlier). Furthermore, We consider GEMM operation between input matrix \textbf{\textit{I}} (with dimensions \textit{C}$\times$\textit{K}=4$\times$4) and weight matrix \textbf{\textit{W}} (with dimensions \textit{K}$\times$\textit{D}=4$\times$4) to generate output matrix \textbf{\textit{O}} (with dimensions \textit{C}$\times$\textit{D}=4$\times$4).

\subsection{Output Stationary Dataflow}
\begin{figure}[h]
  \centering
  \includegraphics[width=\linewidth]{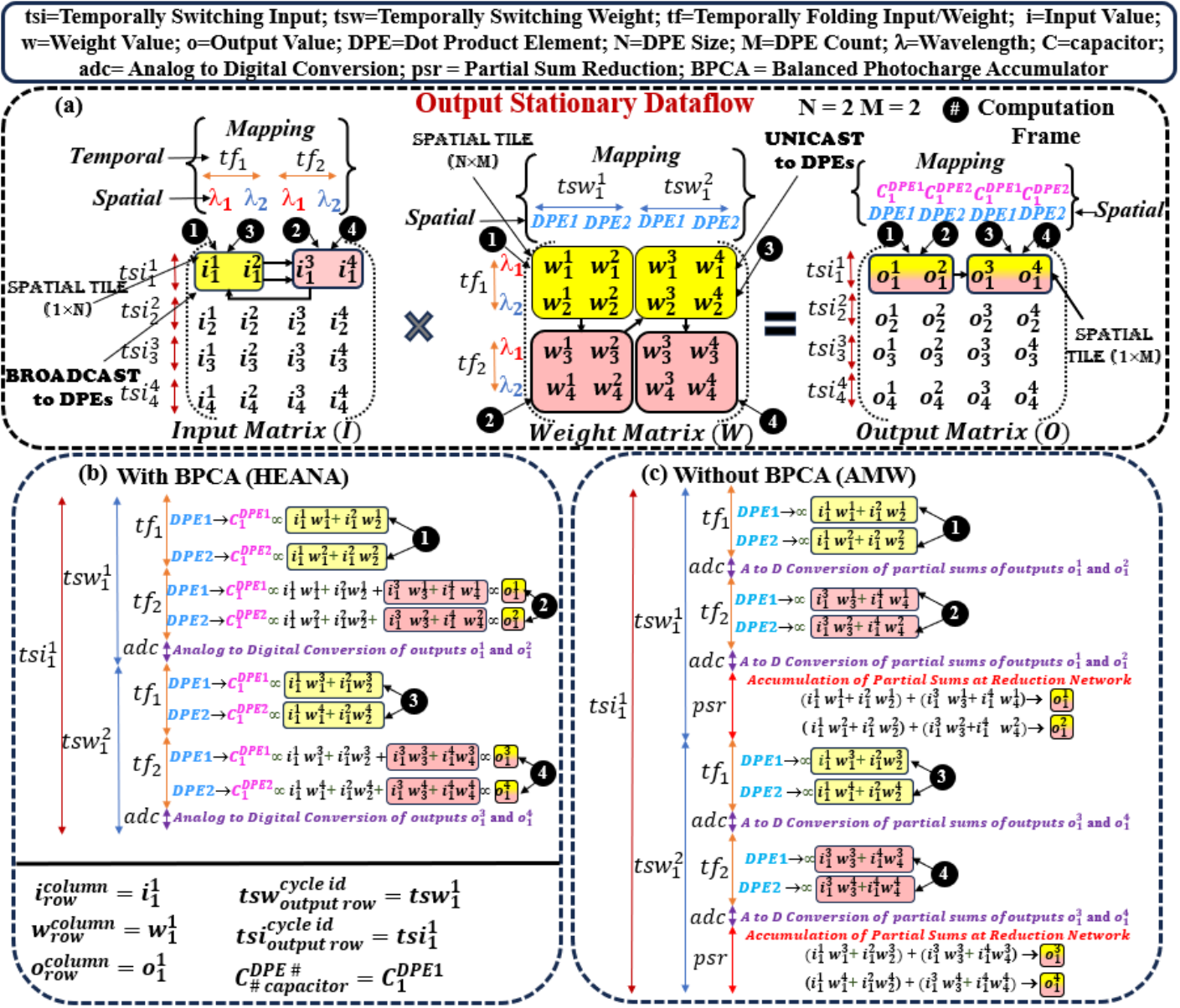}
  \caption{Mapping of GEMM operation between \textbf{\textit{I}} and \textbf{\textit{W}} with output stationary dataflow on HEANA DPU compared to AMW DPU. (a) Temporal and Spatial tiling of \textbf{\textit{I}}, \textbf{\textit{W}}, and \textbf{\textit{O}} depending on DPU parameters \textbf{M} and \textbf{N} (b) Evaluation of output row 1 ($0_1$) by HEANA DPU with BPCA (c) Evaluation of output row 1 ($0_1$) by AMW DPU without BPCA.} 
  \label{outputstationarydataflow}
\end{figure}

Fig. \ref{outputstationarydataflow}(a) illustrates the \textit{OS} dataflow mapping. The tiling of the inputs and weights allows the computing of row 1 (\textbf{\textit{O}}$_1$) of the matrix \textit{\textbf{O}}. Evaluating row 1 of the matrix \textit{\textbf{O}} involves dot product operations between row 1 of the matrix \textit{\textbf{I}} and all the columns of the matrix \textit{\textbf{W}}. These dot products are broken down into multiple computation frames for scheduling/mapping on the DPU. The scheduling order of the computation frames is defined in Fig. \ref{loopunrolling}(a). To map the computation frames, each DPU offers spatial parallelism at two levels, in general, one at the wavelength level and the other at the DPE level. The parameter \textit{N} governs the wavelength-level parallelism, while \textit{M} governs the DPE-level parallelism. For the input matrix \textit{\textbf{I}}, the \textit{OS} dataflow exploits the wavelength-level parallelism in the DPEs by tiling the inputs according to \textit{N = 2} so that the input tile of shape 1$\times$N=1$\times$2 is spatially mapped on the wavelengths $\lambda_1$ and $\lambda_2$ of the DPEs. The same input tile is broadcasted to both the DPEs for spatial reuse of the input values. For the weight matrix \textit{\textbf{W}}, both the wavelength-level and DPE-level parallelisms are leveraged, and according to the parameters \textit{N=2} and \textit{M=2}, the shape of the weight tile is determined as N$\times$M=2$\times$2. Consequently, weights are spatially mapped to wavelengths as well as DPEs. The weight tile is unicast to two DPEs, with each DPE receiving different weight values while using the shared input values to compute different output values. To compute the output row 1 (\textbf{\textit{O}}$_1$), the input and weight tiles undergo temporal switching and folding as shown in Fig. \ref{outputstationarydataflow}(a), to evaluate computation frames (\filledcircled{1},\filledcircled{2},\filledcircled{3}, and \filledcircled{4}). According to the computation order of \textit{OS} dataflow, the mapped output tiles are changed fewer times compared to the input and weight tiles. 

To understand this mapping, consider the example illustrated in Fig. \ref{outputstationarydataflow}(a). In the figure, for computation frame \filledcircled{1}, an input tile (yellow tile) is broadcasted whereas the weight tile (yellow tile) is unicasted to DPEs. Therefore, DPE1 and DPE2 both receive inputs ($i_1^1$, $i_1^2$) while DPE1 receives weights ($w_1^1$,$w_2^1$) and DPE2 receives weights ($w_1^2$,$w_2^2$). For computation frame \filledcircled{1}, DPE1 and DPE2 compute partial sum (\textit{psum}) results towards output values $o_{1}^1$ and $o_{1}^2$, respectively. Then, computation frame changes from \filledcircled{1}$\rightarrow$\filledcircled{2}. For computation frame \filledcircled{2}, the temporal folding ($tf_1$$\rightarrow$$tf_2$) of input and weight tiles from yellow to pink tiles takes place as shown in Fig. \ref{outputstationarydataflow}(a). The updated input and weight tiles (pink tiles) are broadcasted and unicasted to DPEs, respectively. Subsequently, the DPE1 and DPE2 generate the remaining \textit{psum} results ($i_1^3w_3^1$ + $i_1^4w_4^1$) and ($i_1^3w_3^2$ + $i_1^4w_4^2$) towards output values $o_{1}^1$ and $o_{1}^2$, respectively. Thus, the output tile being computed remains constant while the input and weight tiles are changed to evaluate the \textit{psum} results of $o_{1}^1$ and $o_{1}^2$. To obtain the final values of $o_{1}^1$ and $o_{1}^2$, there is a need to accumulate the \textit{psum} results generated in DPE1 and DPE2 via computation frames \filledcircled{1} and \filledcircled{2}. 

After accumulation, the temporal switching of weights ($tsw_1^1$$\rightarrow$$tsw_1^2$) along with the resetting of the temporal folding to $tf_1$ is carried out to implement the computation frame \filledcircled{3}. This switching of weights moves the weight tile to columns 3 and 4 of the matrix \textit{W}, thus resulting in the changing of the output tile to $o_{1}^3$ and $o_{1}^4$. Computation frame \filledcircled{3} evaluates \textit{psum} results for the newly changed output tile. After completion of computation frame \filledcircled{3}, the temporal folding ($tf_1$$\rightarrow$$tf_2$) of input and weight tiles from yellow to pink tiles is again carried out to evaluate computation frame \filledcircled{4}. Computation frame \filledcircled{4} generates the residual \textit{psum} results for $o_{1}^3$ and $o_{1}^4$. The accumulation of \textit{psum} results from computation frame \filledcircled{3} and computation frame \filledcircled{4} generates the final values of $o_{1}^3$ and $o_{1}^4$. This accumulation is performed in HEANA temporally by leveraging the temporal accumulation capability of the BPCA-based design (as explained in Sections 3.2.4 and 3.2.5). With this accumulation, all the values of output row 1 are evaluated. In Fig. \ref{outputstationarydataflow}(a), the arrows connecting the tiles illustrate the direction of tile switching (changing), and the count of arrows between tiles signifies the number of switchings (changes). Thus, from Fig. \ref{outputstationarydataflow}(a), it is evident that for the \textit{OS} dataflow the output tiles are switched (changed) the least number of times compared to the input or weight tiles.

Here, it's important to note that the \textit{psum} results generated by DPEs for evaluating computation frames are in analog form, requiring subsequent analog-to-digital conversion. HEANA and prior optical accelerators differ in the required number of analog-to-digital conversions and the nature of the involved \textit{psum} reduction (\textit{psum} accumulation) process. This is discussed next.

Fig. \ref{outputstationarydataflow}(b) and Fig. \ref{outputstationarydataflow}(c) illustrate the processing of output row 1 \textbf{$O_{1}$} on HEANA and AMW architectures, respectively.
As described in Section \ref{325}, HEANA is integrated with BPCA consisting of capacitors that enable spatio-temporal in-situ accumulations. During temporal switching cycle $tsi_1$, two temporal switching cycles of weights ($tsw_1^1$ and $tsw_1^2$) take place. Within $tsw_1^1$ cyle, the DPU calculates output values $o_{1}^1$ and $o_{1}^2$ involving computation frames \filledcircled{1} and \filledcircled{2}. As explained above, for the computation frame \filledcircled{1}, during $tf_1$, DPE1 and DPE2 generate analog \textit{psums} ($i_1^1w_1^1$+$i_1^2w_2^1$) and ($i_1^1w_1^2$+$i_1^2w_2^2$) towards $o_{1}^1$ and $o_{1}^2$, respectively. In HEANA, analog \textit{psum} values are stored in analog format as proportional voltage levels on capacitors $C_1^{DPE1}$ and $C_1^{DPE2}$ of BPCAs. Subsequently, for frame \filledcircled{2}, during $tf_2$ cycle, weight values ($w_3^1$, $w_4^1$) are mapped to DPE1, and weight values ($w_3^2$, $w_4^2$) are mapped to DPE2. Both DPE1 and DPE2 receive input values ($i_3^1$, $i_4^2$) to generate residual \textit{psum} results towards output values $o_{1}^1$ and $o_{1}^2$. HEANA accrues the generated residual \textit{psum} results by incrementing a proportionate analog voltage on capacitors $C_1^{DPE1}$ and $C_1^{DPE2}$ atop the voltage level stored during the $tf_1$ cycle. The final voltage accrued on capacitors $C_1^{DPE1}$ and $C_1^{DPE2}$ is proportional to the final output values $o_{1}^1$ and $o_{1}^2$. Thus, HEANA employs in-situ temporal accumulation of \textit{psums}. Finally, the analog to digital conversion of $o_{1}^1$ and $o_{1}^2$ is carried out to produce final results as shown in Fig. \ref{outputstationarydataflow}(b). \textit{In contrast}, the AMW architecture lacks BPCAs, and therefore converts the analog \textit{psum} results generated during $tf_1$ and $tf_2$ to digital format using ADCs, before storing these results in buffers. Then, it employs an electronic reduction network such as S\_Tree \cite{maeri2018}, to accumulate the \textit{psum} results from $tf_1$ and $tf_2$ to generate the final values for outputs $o_{1}^1$ and $o_{1}^2$. Thus, AMW requires more analog-to-digital conversions and needs an external reduction network for performing \textit{psum} accumulations. Similarly, $tsw_1^2$ computes $o_{1}^3$ and $o_{1}^4$ involving \filledcircled{3} and \filledcircled{4}, next $tsi_1$ gets updated to $tsi_2$ for generating output pixels of output row 2 \textbf{$O_2$}.     

In summary, HEANA BPCAs offer several advantages. First, it significantly reduces the required count of analog-to-digital conversions, as \textit{psum} results are intermittently stored in BPCA capacitors. This reduction in analog-to-digital conversion count translates into lower power consumption. Second, HEANA leverages the in-situ temporal accumulation capabilities of BPCA, which eliminates the requirement of a reduction network altogether. This elimination leads to a substantial reduction in \textit{psum} reduction latency and the associated energy consumption.

\subsection{Input Stationary Dataflow}
\begin{figure}[h]
  \centering
  \includegraphics[width=\linewidth]{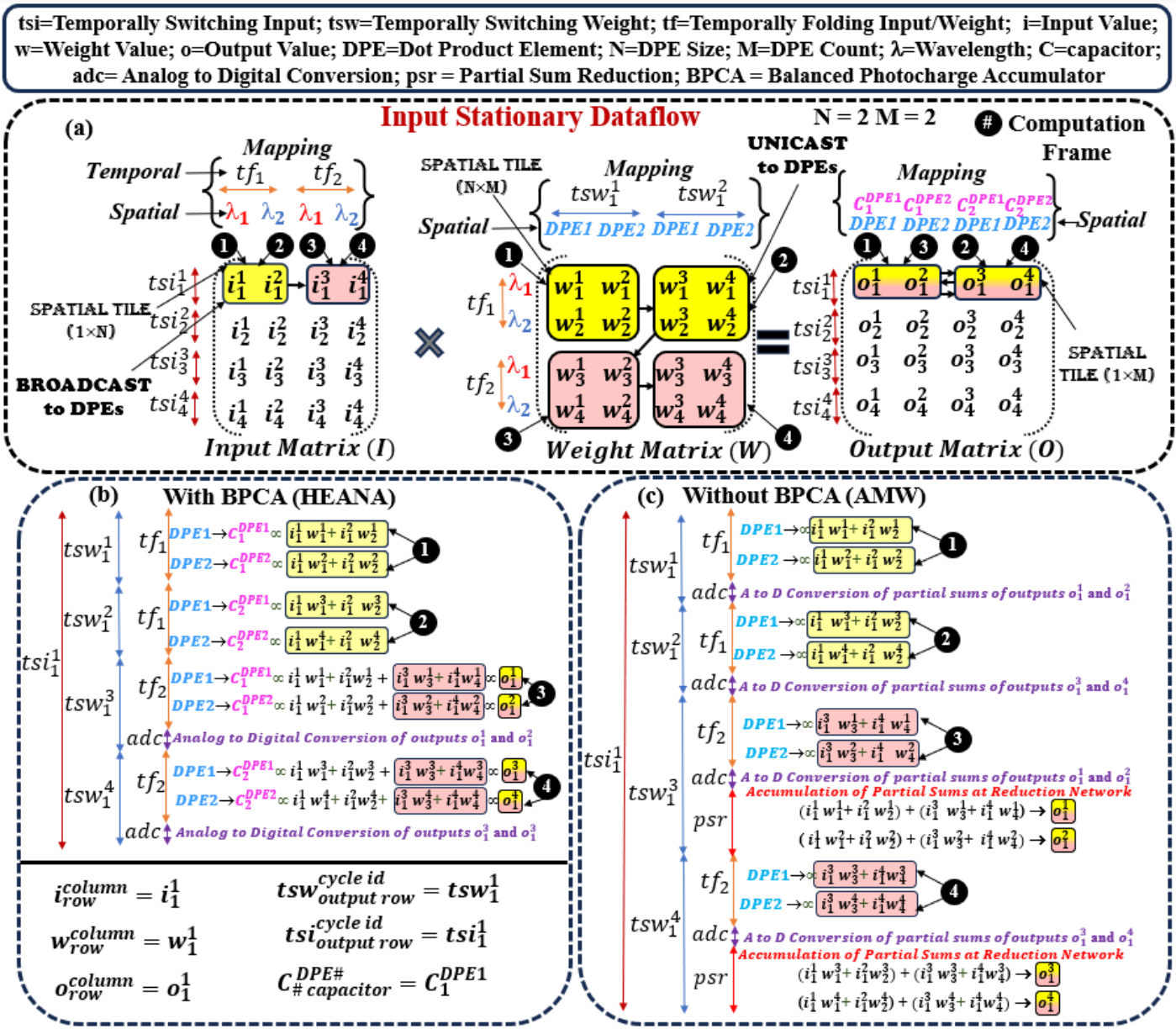}
  \caption{Mapping of GEMM operation between \textbf{\textit{I}} and \textbf{\textit{W}} with input stationary dataflow on HEANA DPU compared to AMW DPU. (a) Temporal and Spatial tiling of \textbf{\textit{I}}, \textbf{\textit{W}}, and \textbf{\textit{O}} depending on DPU parameters \textbf{M} and \textbf{N} (b) Evaluation of output row 1 ($0_1$) by HEANA DPU with BPCA (c) Evaluation of output row 1 ($0_1$) by AMW DPU without BPCA.}  
  \label{inputstationarydataflow}

\end{figure}

Fig. \ref{inputstationarydataflow}(a) illustrates the \textit{IS} dataflow mapping. The input and weight matrices are tiled similar to the tiling for the \textit{OS} dataflow. However, the \textit{IS} dataflow maps with the least number of changes in input tiles. As a result, as shown in Fig. \ref{loopunrolling}(b), the computing functions performed during various computation frames differ between the \textit{IS} and \textit{OS} dataflows. For computation frame \filledcircled{1}, in Fig. \ref{inputstationarydataflow}(a), the input tile (yellow tile) is broadcasted, while the weight tile (yellow tile) is unicasted to DPEs. DPE1 and DPE2 compute the \textit{psum} results toward output pixels $o_{1}^1$ and $o_{1}^2$, respectively. Then, keeping the input tile (yellow tile) unchanged, temporal switching of the weight tile is carried out ($tsw_1^1$$\rightarrow$$tsw_1^2$) to evaluate computation frame \filledcircled{2}, resulting in the generation of \textit{psum} results belonging to $o_1^3$ and $o_1^4$ at DPE1 and DPE2, respectively. After completion of computation frame \filledcircled{2}, all computation frames involving the input tile (yellow tile) are completed. Subsequently, temporal switching of weight tile is again carried out ($tsw_1^2$$\rightarrow$$tsw_1^3$) (labels are not depicted in Fig. \ref{inputstationarydataflow}(a)) along with temporal folding ($tf_1$$\rightarrow$$tf_2$) to perform computation frame \filledcircled{3} for generating the residual \textit{psum} results belonging to output values $o_{1}^1$ and $o_{1}^2$. Finally, temporal switching of weight tile is carried out ($tsw_1^3$$\rightarrow$$tsw_1^4$)(labels are not depicted in Fig. \ref{inputstationarydataflow}(a)) to compute computation frame \filledcircled{4}, generating residual \textit{psum} results belonging to output pixels $o_1^3$ and $o_1^4$. In the \textit{IS} dataflow, as evident from the number of arrows between the tiles shown in Fig. \ref{inputstationarydataflow}(a), the weight and output tiles are changed more often than the input tiles.

Fig.\ref{inputstationarydataflow}(b) and Fig.\ref{inputstationarydataflow}(c) illustrate the processing of output row 1 ($O_{1}$) on HEANA and AMW architectures, respectively. In this \textit{IS} dataflow, the order of temporal weight switching cycle ($tsw$) and temporal folding cycle ($tf$) differ from the \textit{OS} dataflow. Initially, $tsw_1^1$ and $tf_1$ cycles are scheduled to perform computations of frame \filledcircled{1} at DPE1 and DPE2,  generating \textit{psum} results toward $o_1^1$ and $o_1^2$ which involve inputs $i_1^1$ and $i_1^2$. This \ textit{psum} results are generated in the analog format and they are stored as the accrued analog voltage levels on the capacitors $C_1^{DPE1}$ and $C_1^{DPE2}$ using the BPCAs, as from Fig. \ref{inputstationarydataflow}(b). Then, by keeping $i_1^1$ and $i_1^2$ stationary, a temporal switching of weights to cycle $tsw_1^2$ is carried to schedule computation frame \filledcircled{2} to generate \textit{psum} results belonging to $o_1^3$ and $o_1^4$. In this case, HEANA stores the generated \textit{psum} results in different capacitors, namely $C_2^{DPE1}$  and $C_2^{DPE2}$, within the BPCAs. This selection of a different set of capacitors is necessary as the \textit{psum} results generated during cycle $tsw_1^1$ for frame \filledcircled{1} and cycle $tsw_1^2$ for frame \filledcircled{2} belong to different output tiles, and therefore they cannot be accumulated on the same capacitors. This is the reason why our BPCAs incorporate multiple feedback capacitors; multiple capacitors enable the spatio-temporal accumulation and storage of multiple \textit{psum} results corresponding to different output tiles to meet the requirement of different dataflow.      

During cycles $tsw_1^2$ and $tf_1$, all computations involving $i_1^1$ and $i_1^2$ are completed. Following this, cycle $tsw_1^3$ is scheduled in conjunction with cycle $tf_2$, aimed at mapping inputs $i_1^3$ and $i_1^4$ for the computation of frame \filledcircled{3} which generates residual \textit{psum} results associated with $o_1^1$ and $o_1^2$. In this process, HEANA utilizes capacitors $C_1^{DPE1}$ and $C_1^{DPE2}$ to store \textit{psum} results generated during cycles $tsw_1^1$ and $tf_1$. The generated \textit{psum} results for frame \filledcircled{3} are temporally accumulated on top of the \textit{psum} results stored on the capacitors during frame \filledcircled{1}. At the end, the total voltage accrual provides final output values $o_1^1$ and $o_1^2$, which are converted using analog-to-digital conversion. In contrast to HEANA, in Fig. \ref{inputstationarydataflow}(c), the AMW architecture needs multiple analog-to-digital conversions and the utilization of a reduction network to reduce the \textit{psum}s associated with $o_1^1$ and $o_1^2$.

Likewise, cycles $tsw_1^4$ and $tf_2$ are utilized to compute frame \filledcircled{4} for generating $o_1^3$ and $o_1^4$. For that, HEANA utilizes capacitors $C_2^{DPE1}$  and $C_2^{DPE2}$ for in-situ temporal accumulations of \textit{psum}s, whereas AMW once again relies on a reduction network for the accumulation of \textit{psum}s.

\subsection{Weight Stationary Dataflow}
\begin{figure}[h]
  \centering
  \includegraphics[width=\linewidth]{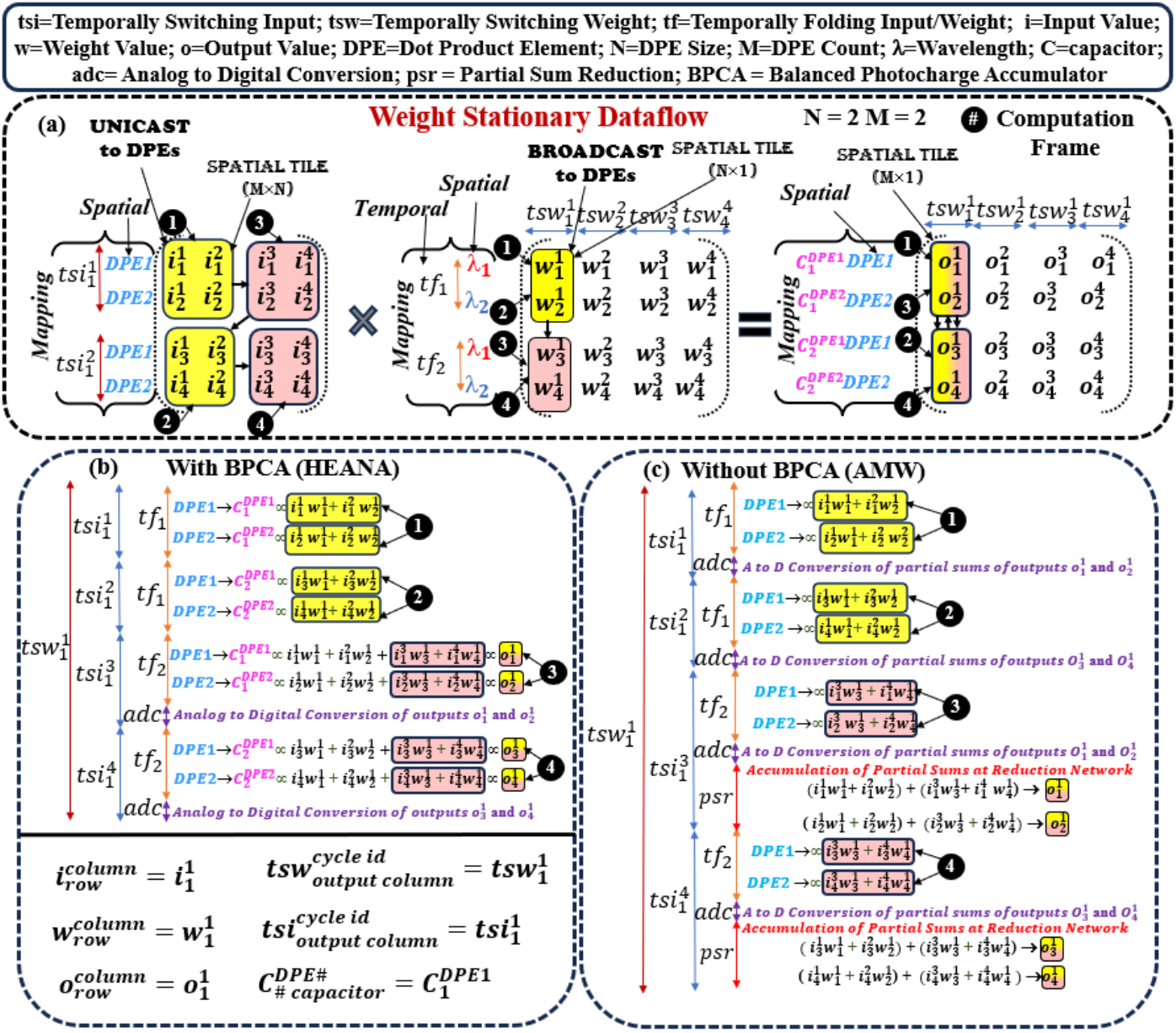}
  \caption{Mapping of GEMM operation between \textbf{\textit{I}} and \textbf{\textit{W}} with weight stationary dataflow on HEANA DPU compared to AMW DPU. (a) Temporal and Spatial tiling of \textbf{\textit{I}}, \textbf{\textit{W}}, and \textbf{\textit{O}} depending on DPU parameters \textbf{M} and \textbf{N} (b) Evaluation of output column 1 ($0^1$) by HEANA DPU with BPCA (c) Evaluation of output column 1 ($0^1$) by AMW DPU without BPCA.} 
  \label{weightstationarydataflow}

\end{figure}

Fig. \ref{weightstationarydataflow}(a) depicts the \textit{WS} dataflow mapping and tiling while computing column 1 (\textbf{\textit{O}}$^1$) of output matrix \textit{\textbf{O}}. Unlike the \textit{IS} and \textit{OS} tiling, in the \textit{WS} dataflow, the inputs are spatially mapped onto both wavelengths and DPEs, while weights are spatially mapped to wavelengths and reused across DPEs. As shown in Fig. \ref{weightstationarydataflow}, the shape of the weight tile depends on \textit{N}, and the input tile shape is determined based on \textit{N} and \textit{M}. In the \textit{WS} dataflow, the required counts of temporal switching for input and output tiles are more compared to the required count of switching of weight tiles. Therefore, the computation order of the tiles in the \textit{WS} dataflow differs when compared to the \textit{IS} and \textit{OS} dataflows. In the \textit{WS} dataflow, for computation frame \filledcircled{1} (see Fig. \ref{weightstationarydataflow}(a)), the input tile (yellow tile) is unicasted, while the weight tile (yellow tile) is broadcasted to DPEs. While processing computation frame \filledcircled{1}, DPE1 and DPE2 compute the \textit{psum} results belonging to output values $o_{1}^1$ and $o_{2}^1$, respectively. Next, to process computation frame \filledcircled{2},  temporal switching of input tile is carried out ($tsi_1^1$$\rightarrow$$tsi_1^2$) to evaluate \textit{psum} results belonging to $o_{3}^1$ and $o_{4}^1$ at DPE1 and DPE2, respectively. During frame \filledcircled{2}, all computing functions involving weight tile (yellow tile) are completed. Subsequently, temporal switching of the input tile is again carried out ($tsi_1^2$$\rightarrow$$tsi_1^3$) (labels are not depicted in Fig. \ref{weightstationarydataflow}(a)) along with temporal folding ($tf_1$$\rightarrow$$tf_2$) to change the yellow tiles to the pink tiles. Here, DPE1 and DPE2 process computation frame \filledcircled{3} for generating the residual \textit{psum} results toward output values $o_{1}^1$ and $o_{2}^1$. Finally, temporal switching of input tile is carried out ($tsi_1^3$$\rightarrow$$tsi_1^4$)(labels are not depicted in Fig. \ref{weightstationarydataflow}(a)) to process computation frame \filledcircled{4} to generate residual \textit{psum} results toward output values $o_3^1$ and $o_4^1$. As evident from Fig. \ref{weightstationarydataflow}(a), the input and output tiles are changed more often than the weight tiles in the \textit{WS} dataflow.  

Fig. \ref{weightstationarydataflow}(b) and Fig. \ref{weightstationarydataflow}(c) illustrate the processing of output column 1 $O_{1}$ on HEANA and AMW architectures employing the \textit{WS} dataflow, respectively. The \textit{WS} dataflow uses temporal input switching ($tsi$) and temporal folding ($tf$). The $tsi_1^1$ and $tf1$ cycles are scheduled for processing computation frame \filledcircled{1}, to generate \textit{psum} results toward $o_1^1$ and $o_2^1$. This involves weights $w_1^1$ and $w_2^1$. The generated psum results are stored in analog format on capacitors $C_1^{DPE1}$ and $C_1^{DPE2}$ by HEANA (Fig. \ref{weightstationarydataflow}(c)). Subsequently, by keeping weights $w_1^1$ and $w_2^1$ stationary, a temporal switching to cycle $tsi_1^2$ is executed for computation frame \filledcircled{2} to generate \textit{psum} results toward output values $o_3^1$ and $o_4^1$. To store these \textit{psum} results, capacitors $C_2^{DPE1}$ and $C_2^{DPE2}$ within the BPCAs are used by HEANA. After cycle $tsi_1^2$, all computations involving $w_1^1$ and $w_2^1$ are completed. Then, computation frame \filledcircled{3} is executed with cycles $tsi_1^3$ and $tf_2$ by mapping weights $w_3^1$ and $w_4^1$, to generate the residual \textit{psum} results associated with output values $o_1^1$ and $o_2^1$. HEANA performs in-situ temporal accumulation of \textit{psum} results generated during computation frames \filledcircled{1} and \filledcircled{3} by using capacitors $C_1^{DPE1}$ and $C_1^{DPE2}$. This accumulation process results in the final voltage levels on the capacitors that are proportional to the final output values $o_1^1$ and $o_2^1$. On the other hand, from Fig. \ref{weightstationarydataflow}(c), AMW architecture derives final output values $o_1^1$ and $o_2^1$ using multiple analog-to-digital conversions and partial sum reduction at the reduction network. Similarly, to map computation frame \filledcircled{4}, the combination of cycles $tsi_1^4$ and $tf_2$ is utilized to generate output values $o_3^1$ and $o_4^1$. Here, HEANA utilizes capacitors $C_2^{DPE1}$ and $C_2^{DPE2}$ for in-situ temporal accumulations of \textit{psum} results corresponding to $o_3^1$ and $o_4^1$, whereas AMW once again relies on a reduction network for the accumulation of \textit{psum} results.  

As evident from the examples shown in Fig. \ref{outputstationarydataflow}, Fig. \ref{inputstationarydataflow}, and Fig. \ref{weightstationarydataflow}, HEANA requires fewer analog-to-digital conversions to generate an output value. Table \ref{table1} reports the number of analog-to-digital conversions needed by a DPU to compute all the output values of a GEMM operation. From the table, the AMW and MAW architectrues require $ceil(K/N)\times$ more conversions than HEANA. Hence, HEANA can significantly reduce the energy and latency of analog-to-digital conversions as shown in our system-level analysis (see Section \ref{section6}). Furthermore, it is worth noting that HEANA is not only more efficient in terms of energy consumption and latency but also more flexible and compatible with a wide range of input dimensions.

\begin{table}[]

\caption{Number of analog-to-digital (AtoD) conversions required by various optical DPUs for a GEMM operation. C and D are the height and width of the output matrix, where K is the width (height) of the input (weight) matrix. N=DPE size.}
\label{table1}     
\begin{tabular}{|c|c|c|}
\hline
\textbf{DPU}             & \textbf{Dataflow} & \textbf{\begin{tabular}[c]{@{}c@{}}AtoD\\ Conversions\end{tabular}} \\ \hline
\multirow{3}{*}{HEANA}   & OS                & C$\times$D                                                                  \\ \cline{2-3} 
                         & WS                & C$\times$D                                                                  \\ \cline{2-3} 
                         & IS                & C$\times$D                                                                  \\ \hline
\multirow{3}{*}{AMW/MAW} & OS                & C$\times$D$\times$ceil(K/N)                                                        \\ \cline{2-3} 
                         & WS                & C$\times$D$\times$ceil(K/N)                                                        \\ \cline{2-3} 
                         & IS                & C$\times$D$\times$ceil(K/N)                                                        \\ \hline
\end{tabular}
\end{table}

\subsection{Operational Logistics and Benefits of HEANA} 
\underline{Capacitor Selection Based on Dataflows}: For the \textit{OS} dataflow, the \textit{psum} results generated by the BPCA in consecutively scheduled computation frames belong to the same output value. Therefore, the same capacitor is selected for several continuous frames until the accumulation of all the \textit{psum} results belonging to the same output value has been completed (Fig. \ref{outputstationarydataflow}(b)). In contrast, for the \textit{IS} dataflow (Fig. \ref{inputstationarydataflow}(b)) and the \textit{WS} dataflow (Fig. \ref{weightstationarydataflow}(b)), the \textit{psum} results generated by the BPCA in consecutive frames belong to different output values. Therefore, a different capacitor is selected for every frame by controlling the corresponding switches (Fig. \ref{HEANA}) using a demux. From the sizes of the Toeplitz matrices of the state-of-the-art CNNs \cite{chollet2015keras}, we have determined that the BPCA in our HEANA architecture may require up to \textit{p=4608} capacitors to support seamless accumulation of \textit{psum} results during the execution of the \textit{IS} and \textit{WS} dataflow. Furthermore, our evaluation revealed  the latency and power corresponding to the switching of capacitors for the IS and WS dataflows as 2.5 ns and  0.041 mW, respectively \cite{dsent}. Furthermore, it is noteworthy that the overhead of adding multiple capacitors to the baseline time-integrating receiver circuit \cite{alexandermit2022,BhaskaranPCA2022} to build our proposed BPCA is 0.016 $mm^2$, which is very low.

\underline{Benefits due to the optical domain operation of HEANA:} There is a two-pronged benefit: (i) a massive fan-in of signals for massive spatial parallelism, and (ii) the summation of a massive number of optical symbols spatially and temporally using a single electro-photonic adder based on our proposed BPCA design. This benefit alludes electronic accelerators even if they employ time-amplitude analog modulator-based architecture for multiplications instead of the traditional amplitude-amplitude analog modulation-based architecture. 

\underline{Accumulation Benefits of HEANA}\label{benefits}: In HEANA, a \textit{psum} result is accumulated as a voltage level accrued on a capacitor of a BPCA. After each frame, the capacitor can hold on to the accrued voltage for a significant amount of time, thus eliminating the need to convert the \textit{psum} result in the digital format via ADC and store it in a buffer temporarily. This eliminates the latency and energy overheads related to ADC conversions and digital buffering (read+write) of \textit{psum} results. Several of such \textit{psum} results might be required to be accumulated together to produce one value in the output matrix \textbf{\textit{O}} (Fig. \ref{fig1}). The use of TIR in our BPCA enables the accumulation of several such \textit{psum} results temporally over multiple frames (over continuously, intermittently, or sporadically scheduled frames, depending on the utilized dataflow), by allowing a linear increment of the accrued voltage on the capacitor proportion to the psum result arriving at each frame. This eliminates the need to employ dedicated \textit{psum} reduction networks, thereby eliminating related latency and energy overheads as well.

Furthermore, due to the incoherent superposition of optical pulses at BPDs, the BPDs accumulate all incident optical pulses arriving during a period equivalent to the inverse bandwidth of the PD to provide an output photo-current whose amplitude is proportional to the sum of optical power of all of the incident optical pulses \cite{}. For instance, a PD operating at 1 GHz with the inverse bandwidth of 1 ns can accumulate 10 optical pulses arriving with a 1 ns period at the datarate of 10 GHz. Subsequently, an operational amplifier (op-amp) based time-integrating receiver can sample the PD output current and integrate it over the sampling period (equal to the inverse bandwidth of the PD) to accrue a corresponding analog voltage across the capacitor. Therefore, in our HEANA, even if TAOMs operate at 10 GS/s, the maximum operational data rate of the TIR is constrained to 1 GHz. This allows the TIR to run at a significantly lower data rate than the TAOMs, which results in lower power consumption and area usage, contributing to the overall efficiency of the HEANA system.


\section{Scalability Analysis}\label{section5}
\begin{figure}[h]
  \centering
  \includegraphics[scale=0.6]{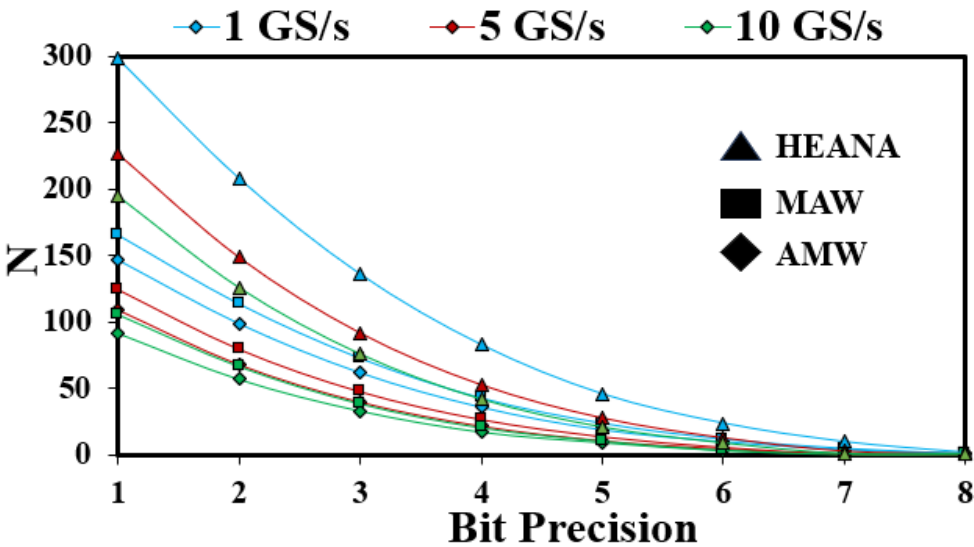}
  \caption{Supported DPU size N (=M) for bit precision=\{1, 2, 3, 4, 5, 6, 7, 8\} bits at data rates (DRs)=\{1, 5, 10\} GS/s, for AMW, MAW, and HEANA DPUs.} 
  \label{fig6}
\end{figure}

To determine the achievable size \textit{N} for our HEANA DPU, we adopt the scalability analysis equations (Eq. \ref{eq3}, Eq. \ref{eq4}, and Eq. \ref{eq5}) from \cite{lukasscalability} and \cite{cases2022,sconna}. Table \ref{abbrevations} reports the definitions of the parameters and their values used in these equations. 
  We consider \textit{M=N} and first solve Eq. \ref{eq3} and Eq. \ref{eq4} for a set of DRs=\{1, 5, 10\} GS/s, to find a corresponding set of $P_{PD-opt}$. Then, we solve Eq. \ref{eq5} for the maximum value of \textit{N} that achieves $P_{O/p}$  greater than obtained $P_{PD-opt}$ values across the set of \textit{DR}s. Fig. \ref{fig6} reports the achievable \textit{N} of HEANA, AMW and MAW DPUs for different bit-precision levels (B) across various \textit{DR}s.  The achievable \textit{N} value defines the feasible number of multipliers per DPE; thus, this \textit{N} also defines the maximum size of the spatial dot product that can be generated in our DPU. As evident from Fig.\ref{fig6}, our HEANA can support larger \textit{N} value compared to AMW and MAW at all bit-precision levels across different DRs. For instance, HEANA achieves larger \textit{N=83} for 4-bit precision at 1 GS/s, compared to AMW and MAW, which achieve \textit{N=36} and \textit{N=43}, respectively. This is due to HEANA's spectrally hitless DPE architecture (as discussed earlier in Section \ref{sec321}) due to which HEANA significantly reduces the crosstalk-related power penalty contributing to $P_{penalty}$. Despite the hitless DPE architecture, inter-channel crosstalk can still occur within the aggregation lanes of HEANA DPEs, particularly at mono-wavelength filters, when dropped signals traverse neighboring filters before reaching the BPD. Based on prior work \cite{karen2020proceeding}, through design choices such as employing mono-wavelength filters with a finesse (F) approximately equal to the size of DPE (N), i.e., F=83, and a quality factor (Q) of around 3000, HEANA effectively minimizes inter-channel crosstalk to approximately 1 dB. This design and architecture choices allow HEANA to support larger \textit{N} compared to AMW and MAW at the same input laser power.




\begin{equation}
       B = \frac{1}{6.02}\Bigg[20log_{10}(\frac{R\times P_{PD-opt}}{\beta\sqrt{\frac{DR}{\sqrt{2}}}})-1.76\Bigg]
       \label{eq3}
\end{equation}

\begin{equation}
    \beta = \sqrt{2q(RP_{PD-opt}+I_d)+\frac{4kT}{R_L}+R^2P_{PD-opt}^2RIN} +\sqrt{2qI_d+\frac{4kT}{R_L}}
    \label{eq4}
\end{equation}
    
\begin{dmath}
\label{eq5}
     P_{O/p}(dBm) = P_{Laser}-P_{SMF-att}-P_{EC-IL}-P_{Si-att}\times N\times d_{MRR}-P_{MRM-IL}-(N-1)P_{MRM-OBL}-P_{splitter-IL}\times log_{2}(M)-P_{MRR-W-IL}-(N-1)P_{MRR-W-OBL}-P_{penalty}-10log_{10}(N)
\end{dmath}


\begin{table}[]
\centering
\caption{Definition and values of various parameters used in Eq. \ref{eq3}, Eq. \ref{eq4}, and Eq. \ref{eq5} (from \cite{lukasscalability,cases2022,sconna}) for the scalability analysis. }
\label{abbrevations}
\begin{tabular}{|c|c|c|}
\hline
{ \textbf{Parameter}}             & { \textbf{Definition}}                                                                                                         & { \textbf{Value}}   \\ \hline
{$P_{Laser}$}                & { Laser Power Intensity}                                                                                               & { 10 dBm}  \\ \hline
 { $R_{s}$}                     & { PD Responsivity}                                                                                                     & { 1.2 A/W} \\ \hline
{ $R_L$}                    & { Load Resistance}                                                                                                     & { 50 $\Omega$}      \\ \hline
{ $I_d$}                    & { Dark Current}                                                                                                        & { 35 nA}   \\ \hline
 { T}                     & { Absolute Temperature}                                                                                                & { 300 K}   
\\ \hline
{ RIN}                   & { Relative Intensity Noise}                                                                                            & { -140 dB/Hz}   \\ \hline 
{ $P_{EC-IL}$}            &\begin{tabular}[c]{@{}c@{}}Fiber to Chip Coupling \\ Insertion Loss\end{tabular}                                                                                & { 1.44 }     \\ \hline
{ $P_{MRR-W-IL}$}         & \begin{tabular}[c]{@{}c@{}} Silicon Waveguide  \\ Insertion Loss\end{tabular}                                                                                   & { 0.3 dB/mm}     \\ \hline
 { $P_{splitter-IL}$}      & { Splitter Insertion Loss}                                                                                             & { 0.01 dB}    \\ \hline
 { $P_{MRM-IL}$}           & \begin{tabular}[c]{@{}c@{}}Optical Microring Modulator  \\ Insertion Loss\end{tabular}                                                                           & { 4 dB}       \\ \hline
  { $P_{MRR-IL}$}           & \begin{tabular}[c]{@{}c@{}}Optical Microring Resonator  \\ Insertion Loss\end{tabular}                                                                           & { 0.01 dB}       \\ \hline
 { $P_{MRM-OBL}$}          & { Out of Band Loss}                                                                                                & { 0.01 dB}                                                                                 \\ \hline
 \multirow{0}{*}{$P_{Penalty}$} & MAW Network Penalty  & 4.8 dB\\ \cline{2-3} 
                    & AMW Network Penalty  & 5.8 dB \\ \cline{2-3} 
                    & HEANA Network Penalty & 1.8 dB\\ \hline

\end{tabular}
\end{table}

\section{Evaluation}
\begin{figure}[h]
  \centering
  \includegraphics[scale=0.7]{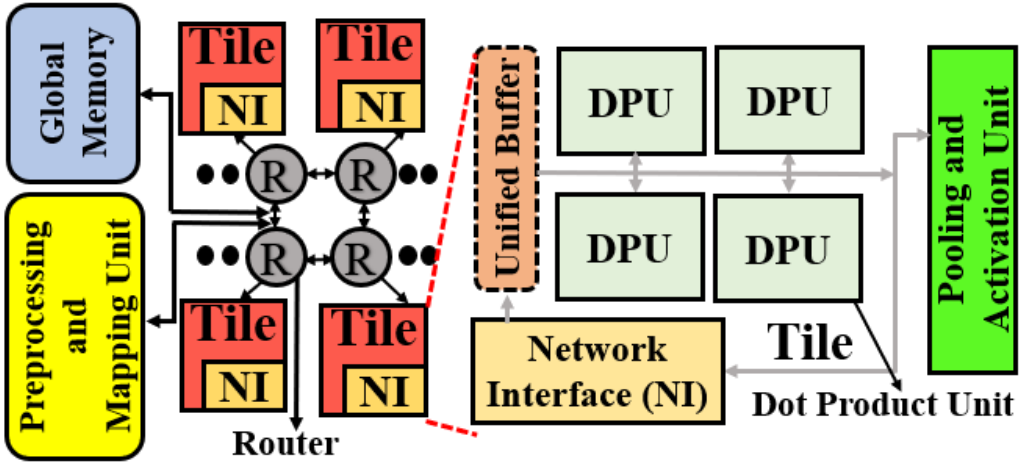}
  \caption{System-level overview of our HEANA accelerator.} 
  \label{fig7}
\end{figure}

\subsection{System Level Implementation of HEANA}
Fig. \ref{fig7} illustrates the system-level implementation of our HEANA accelerator. It consists of global memory that stores CNN parameters and a preprocessing and mapping unit. It has a mesh network of tiles. Each tile contains 4 DPUs interconnected (via H-tree) with a unified buffer, as well as pooling and activation units. Each DPU consists of multiple DPEs and each DPE is equipped with a dedicated input and output FIFO buffer \cite{Wang2021-ISCAS} to store intermittent weights, inputs, and \textit{ psum} results. 

\subsection{Simulation Setup}
For evaluation, we modeled our HEANA accelerator from Fig. \ref{fig7} using our custom-developed, transaction-level, event-driven, python-based simulator (more on this in next subsection). We simulated the inference of four CNNs (batch size=1 and batch size=256): GoogleNet \cite{googlenet}, ResNet50 \cite{resnet}, MobileNet\_V2 \cite{mobilenetv2}, and ShuffleNet\_V2 \cite{shufflenet}. We evaluate frames-per-second (FPS) and FPS/W (energy efficiency).

Prior works \cite{cansu2021, deapcnn, crosslight, albireo, holylight, shen2017deep, Zhang22Optica} have demonstrated the superiority of incoherent and coherent optical accelerators over electronic counterparts in terms of throughput and energy efficiency. Therefore, we focused our comparison on optical accelerators and omitted direct comparisons with electrical accelerators. Direct system-level comparisons with prior coherent optical accelerators are challenging due to methodological disparities and reproducibility issues. Moreover, coherent optical accelerators, while offering power-efficient computation, often rely on complex and large-scale components such as Mach-Zehnder interferometers (MZIs). This introduces significant area overheads and higher power consumption \cite{harris2014efficient}, potentially leading to lower performance compared to incoherent optical accelerators. Hence, we compared our HEANA with the prior incoherent MRR-based accelerators AMW \cite{deapcnn} and MAW \cite{holylight} accelerators. Our BPCA can be easily integrated into AMW and MAW accelerators. Therefore, we have considered two variants of AMW and MAW: (1) AMW and MAW (2) AMW integrated with BPCA (AMW$_{BPCA}$) and MAW integrated with BPCA (MAW$_{BPCA}$). Each accelerator variant is evaluated for the weight stationary (\textit{WS}), input stationary (\textit{IS}), and output stationary (\textit{OS}) dataflows. All accelerators are operated for 4-bit integer precision across datarates 1GS/s, 5GS/s, and 10GS/s. From Fig. \ref{fig6}, for these parameters HEANA, AMW, and MAW achieve N reported in Table \ref{table3}. For a fair comparison, we performed area proportionate analysis, wherein we scaled the DPU count for each photonic CNN accelerator so that the total area of DPUs in each accelerator matches with the area of HEANA (\textit{N=83}) having 50 DPUs. Table \ref{table3} reports the scaled DPU count of AMW, MAW, and HEANA at various data rates. Table \ref{table4} gives the parameters used for our evaluation. Next, we discuss the implementation details of our simulator. 

\subsection{Simulator Overview}
\begin{figure*}[h]
  \centering
  \includegraphics[width=0.8\linewidth]{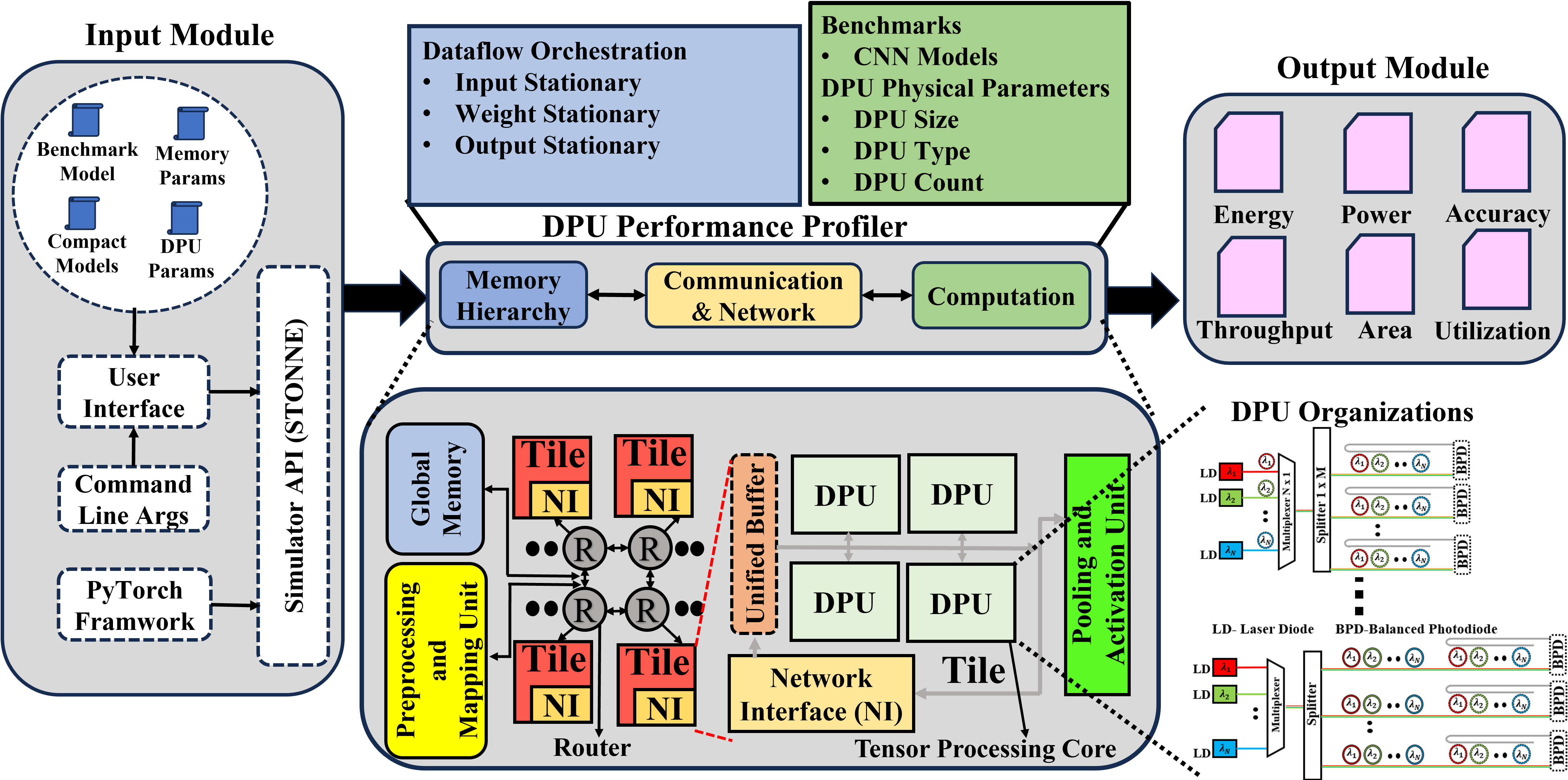}
  \caption{ \textcolor{blue}{High-level diagram of the our in-house simulator.}} 
  \label{simulator}
\end{figure*}

Fig. \ref{simulator} shows the high-level overview of our simulator. Our simulator leverages photonic foundry-validated Ansys Lumerical tools \cite{lumerical_2021} and Multisim \cite{multisim} to model and simulate our TAOM device and BPCA circuit. Using the tools mentioned, we extracted compact models capturing the electro-photonic behavior of the TAOM and BPCA circuits. These models were then imported into the Ansys Lumerical INTERCONNECT tool to enable simulations-based analysis of a TAOM and BPCA-based dot-product element. This process allows us to estimate the performance and energy consumption of the dot-product element with reasonable accuracy.
At the system level, we extended the STONNE simulator \cite{munoz2020stonne} to create a custom transaction-level photonic accelerator simulator using Python and C++. As shown in Fig. \ref{simulator}, the simulator consists of three main modules: the input module, the dot product unit performance profiler, and the output module. The input module configures the simulator based on user-defined settings, including energy models obtained from Ansys Lumerical and Multisim-based simulations. Once configured, the dot product unit performance profiler conducts detailed simulations while gathering performance statistics. Finally, the output module produces files containing execution statistics once the simulation is complete.

\begin{table}[]
\caption{DPU size (N) and DPU Count (\#) at 4-bit precision across various DRs for
different accelerators.}
\label{table3}
\begin{tabular}{|c|cccccc|}
\hline
\multicolumn{1}{|l|}{} & \multicolumn{6}{c|}{\textbf{Datarate}}                                                                                                   \\ \hline
                       & \multicolumn{2}{c|}{\textbf{1 GS/s}}                        & \multicolumn{2}{c|}{\textbf{5 GS/s}}                        & \multicolumn{2}{c|}{\textbf{10 GS/s}}   \\ \hline
\textbf{DPU} &
  \multicolumn{1}{c|}{\textbf{N}} &
  \multicolumn{1}{c|}{\textbf{\#}} &
  \multicolumn{1}{c|}{\textbf{N}} &
  \multicolumn{1}{c|}{\textbf{\#}} &
  \multicolumn{1}{c|}{\textbf{N}} &
  \textbf{\#} \\ \hline
\textbf{AMW}           & \multicolumn{1}{c|}{36} & \multicolumn{1}{c|}{207} & \multicolumn{1}{c|}{17} & \multicolumn{1}{c|}{900} & \multicolumn{1}{c|}{12} & 1950 \\ \hline
\textbf{MAW}           & \multicolumn{1}{c|}{43} & \multicolumn{1}{c|}{280} & \multicolumn{1}{c|}{21} & \multicolumn{1}{c|}{1100} & \multicolumn{1}{c|}{15} & 1610 \\ \hline
\textbf{HEANA}         & \multicolumn{1}{c|}{83} & \multicolumn{1}{c|}{50}  & \multicolumn{1}{c|}{42} & \multicolumn{1}{c|}{180} & \multicolumn{1}{c|}{30} & 320  \\ \hline
\end{tabular}
\end{table}

\begin{table}[]
\begin{threeparttable}[b]
\caption{Accelerator Peripherals and DPU Parameters {\cite{cases2022}}}
\label{table4}
\begin{tabular}{|c|ccc|}
\hline
                           & \multicolumn{1}{c|}{\textbf{Power(mW)}} & \multicolumn{1}{c|}{\textbf{Latency}} & \textbf{Area($mm^2$)} \\ \hline
\textbf{Reduction Network} & \multicolumn{1}{c|}{0.050}              & \multicolumn{1}{c|}{3.125ns}          & 3.00E-5            \\ \hline
\textbf{Activation Unit}   & \multicolumn{1}{c|}{0.52}               & \multicolumn{1}{c|}{0.78ns}           & 6.00E-5            \\ \hline
\textbf{IO Interface}      & \multicolumn{1}{c|}{140.18}             & \multicolumn{1}{c|}{0.78ns}           & 2.44E-2            \\ \hline
\textbf{Pooling Unit}      & \multicolumn{1}{c|}{0.4}                & \multicolumn{1}{c|}{3.125ns}          & 2.40E-4            \\ \hline
\textbf{eDRAM}             & \multicolumn{1}{c|}{41.1}               & \multicolumn{1}{c|}{1.56ns}           & 1.66E-1             \\ \hline
\textbf{Bus}               & \multicolumn{1}{c|}{7}                  & \multicolumn{1}{c|}{5 cycles}         & 9.00E-3               \\ \hline
\textbf{Router}            & \multicolumn{1}{c|}{42}                 & \multicolumn{1}{c|}{2 cycles}         & 1.50E-2              \\ \hline
\textbf{DAC (ALL)  \cite{dac1} }      & \multicolumn{1}{c|}{12.5}         & \multicolumn{1}{c|}{0.78ns}             & 2.50E-3                 \\ \hline
\textbf{DAC(HEANA)\cite{dac10}  }     & \multicolumn{1}{c|}{26}         & \multicolumn{1}{c|}{0.78ns}             &  6.00E-3               \\ \hline
\textbf{\textcolor{blue}{BPCA}\cite{EyangISCAS2019}  }     & \multicolumn{1}{c|}{1.15}         & \multicolumn{1}{c|}{0.78ns}             &  5.2E-3               \\ \hline
\textbf{EO Tuning}         & \multicolumn{1}{c|}{80 $\mu$W/FSR}          & \multicolumn{1}{c|}{20ns}             & -                  \\ \hline
\textbf{TO Tuning}         & \multicolumn{1}{c|}{275 mW/FSR}         & \multicolumn{1}{c|}{4$\mu$s}             & -                  \\ \hline

\end{tabular}

  \end{threeparttable}
\end{table}

\begin{figure*}[h]
  \centering
  \includegraphics[width=\linewidth]{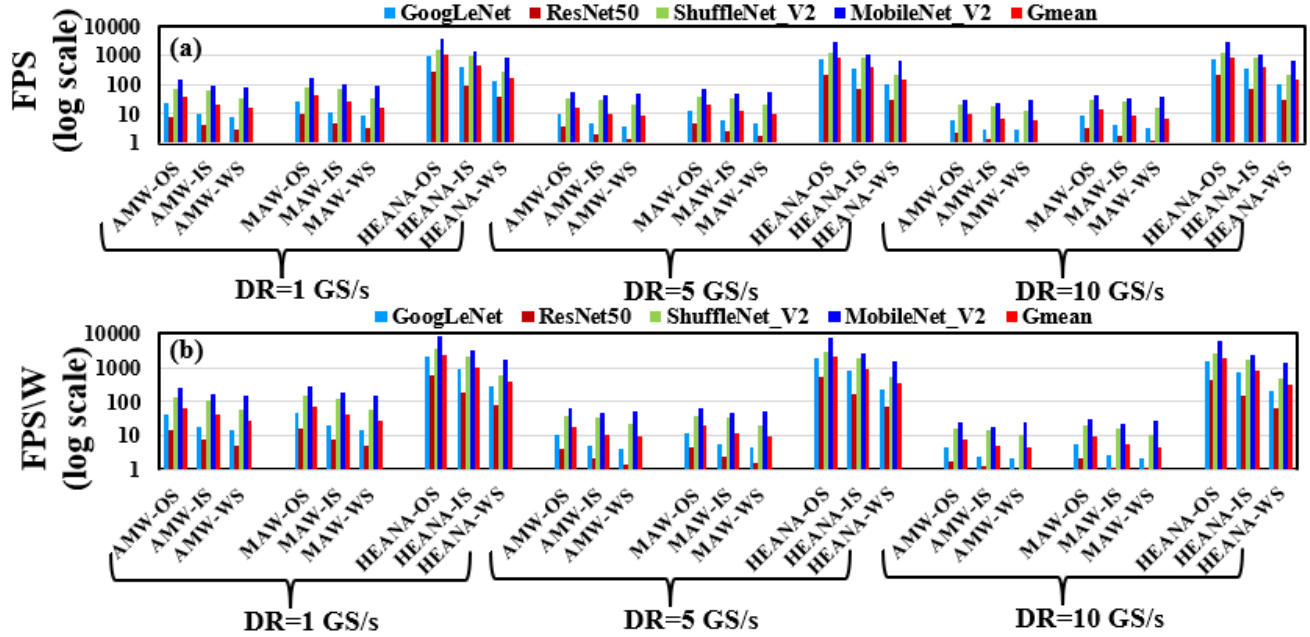}
  \caption{(a) Normalized FPS (log scale) (b) Normalized FPS/W for HEANA versus AMW and MAW accelerators with input batch size=1. Results of FPS and FPS/W are normalized with respect to AMW executing weight stationary dataflow (AMW-WS) for ResNet50 at 10 GS/s.} 
  \label{fig9}
\end{figure*}
\begin{figure*}[h]
  \centering
  \includegraphics[width=\linewidth]{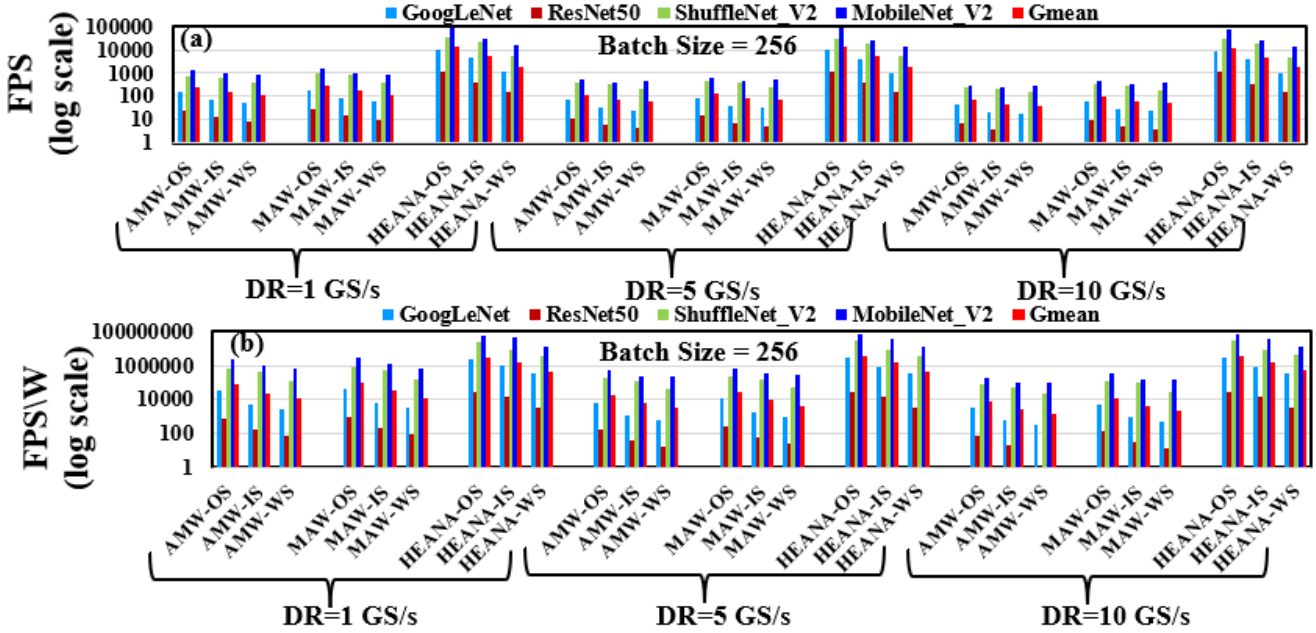}
  \caption{(a) Normalized FPS (log scale) (b) Normalized FPS/W for HEANA versus AMW and MAW accelerators with input batch size=256. FPS and FPS/W results are normalized with respect to AMW executing weight stationary dataflow (AMW-WS) for ResNet50 at 10 GS/s.} 
  \label{fig10}
\end{figure*}

\subsection{Evaluation Results}\label{section6}

Fig. \ref{fig9}(a) shows normalized FPS results for various accelerators with batch size=1 at different datarates (DRs), normalized to AMW-WS for ResNet50 at 10 GS/s. Our accelerator HEANA outperforms MAW and AMW for the \textit{IS}, \textit{OS}, and \textit{WS} dataflows across all data rates. At 1 GS/s, HEANA-OS achieves up to 30$\times$ and 25$\times$ better FPS than AMW and MAW, respectively, across all dataflows. As DR increases to 5 GS/s and 10 GS/s, our HEANA-OS further improves FPS, achieving up to 69$\times$ and 113$\times$ better FPS than AMW, respectively, across all dataflows. HEANA-OS achieves up to 55$\times$ and 83$\times$ better FPS than MAW at 5 GS/s and 10 GS/s, across all dataflows.

These significant improvements in throughput for HEANA are due to two reasons. First, our HEANA with spectrally hitless architecture supports a larger DPU size (\textit{N=83}), i.e., the size of the spatial dot product operation \textit{N} (refer Table \ref{table3}) and the number of parallel dot product operations \textit{M} (=\textit{N}), which increases the overall throughput with improved parallelism.  Second, our BPCA eliminates the latency corresponding to \textit{psum} reductions and corresponding buffer accesses for all dataflows as discussed in Sections \ref{325} and \ref{benefits}. 

Among dataflows, the \textit{ OS} dataflow achieves better throughput for HEANA compared to the \textit{WS} and \textit{IS} dataflows. For HEANA-OS, our BPCA activates the incoherent superposition \cite{BhaskaranPCA2022} at BPD for accumulation (see Section \ref{325}), allowing TAOMs to operate at least 10$\times$ faster than the sample rate of the BPCA. Thus, HEANA-OS achieves FPS of up to 2.3$\times$ and 6.2$\times$ better than HEANA-IS and HEANA-WS, respectively, across all DRs.

In the case of AMW and MAW architectures, the \textit{OS} dataflow outperforms the \textit{WS} and \textit{IS} dataflow. The \textit{OS} dataflow requires the fewest number of output buffer accesses compared to \textit{IS} and \textit{WS}. Additionally, the \textit{OS} dataflow allows for the temporal accumulation of \textit{psums} at a reduction network; therefore, consecutively arriving \textit{psums} can be added together using a temporal accumulator \cite{maeri2018} without having to store them temporarily in a buffer. Therefore, for AMW and MAW, the \textit{OS} dataflow is better than the \textit{IS} and \textit{WS} dataflows. Furthermore, the WS dataflow achieves a lower FPS than the \textit{IS} dataflow for AMW and MAW, as it incurs a higher buffer access latency due to more frequent buffer accesses.

Fig. \ref{fig9}(b) shows FPS/W (log scale) results for various accelerators with batch size=1 at different DRs, normalized to AMW-WS for ResNet50 at 10 GS/s. Our HEANA accelerator on gmean achieves better FPS / W across four CNNs and outperforms MAW and AMW for \textit{IS}, \textit{OS}, and \textit{WS} dataflows across all datarates. At 1 GS/s, HEANA-OS on gmean achieves up to 36$\times$ and 32$\times$ better FPS/W than AMW and  MAW, respectively, across all the dataflows. Similarly, at 5 GS/s and 10 GS/s, our HEANA-OS achieves up to 120$\times$ and 244$\times$ better FPS/W compared to AMW, across all dataflows. HEANA-OS achieves up to 104$\times$ and 204$\times$ better FPS/W compared to MAW at 5 GS/s and 10 GS/s, across all dataflows. HEANA-IS and HEANA-WS also achieve up to 137$\times$ and 54$\times$ better FPS/W compared to AMW and MAW across all the datarates. Our HEANA consumes less energy and static power compared to AMW and MAW. In HEANA, energy consumption related to \textit{psum} buffer accesses is significantly reduced because of BPCA's in-situ spatio-temporal accumulations. The use of BPCA in HEANA also eliminates the energy consumption of the external \textit{psum} reduction network. As discussed in section \ref{section4}, HEANA requires fewer ADC conversions, contributing to energy savings. Additionally, while AMW and MAW employ two MRRs to perform a single multiplication, HEANA employs a single MRR-based TAOM, thereby reducing static power consumption. These benefits collectively result in better FPS/W for HEANA-OS, HEANA-WS, and HEANA-IS. HEANA-OS achieves at least 2.1$\times$ and 6$\times$ better FPS/W than HEANA-WS and HEANA-IS across all the datarates. As DR increases, the FPS/W decreases across all the architectures due to increased energy consumption of ADCs and DACs \cite{cases2022}. 

Fig. \ref{fig10}(a) and Fig. \ref{fig10}(b) shows FPS (log scale) and FPS/W (log scale) results for various accelerators with batch size=256 at different DRs, normalized to AMW-WS for ResNet50 at 10 GS/s. Larger batch sizes yield significantly better benefits from HEANA, as HEANA-OS on gmean achieves up to 347$\times$ better FPS than other architectures across all datarates. Similarly, as the batch size grows, the advantages of HEANA become more pronounced in FPS/W results. HEANA-OS on gmean achieves up to 952$\times$ better FPS/W than other architectures across all the datarates. We found that the impact of dataflow choice is similar to that of batch size=1. Next we discuss the results of HEANA when MAW and AMW are integrated with BPCA.

\begin{figure*}[h]
                \centering
  \includegraphics[width=\linewidth]{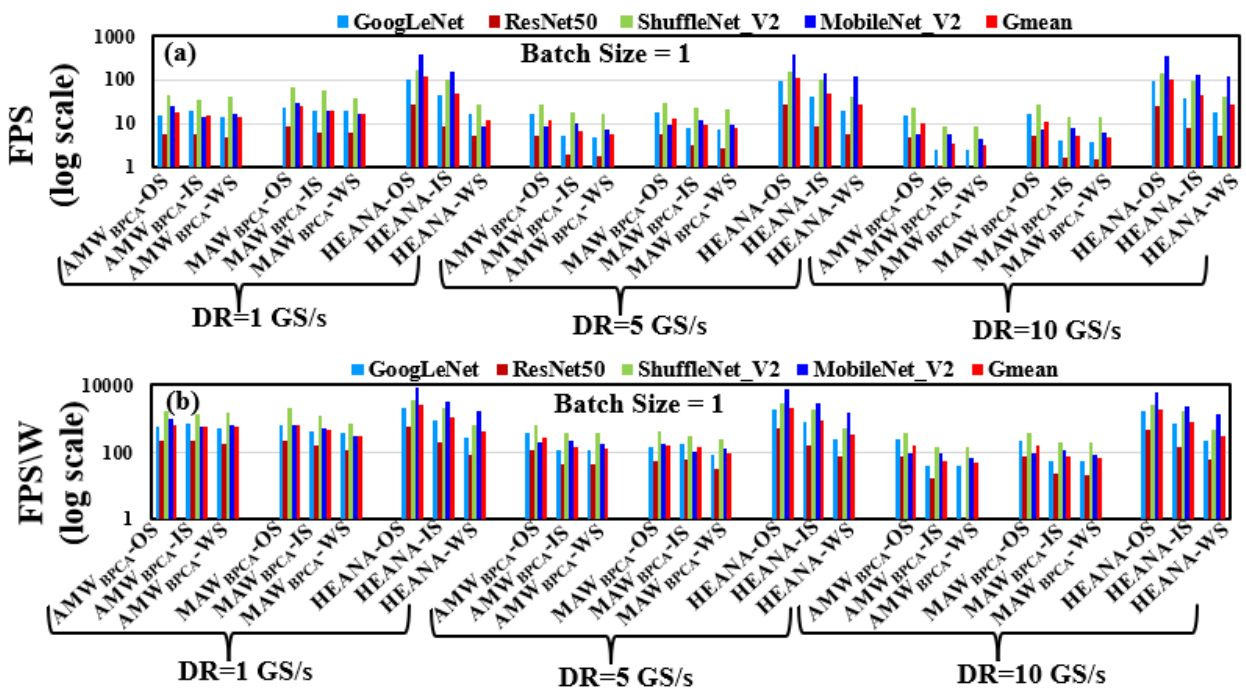}
  \caption{(a) Normalized FPS (log scale), (b) Normalized FPS/W for HEANA versus BPCA-integrated versions of AMW and MAW accelerators with input batch size=1. Results of FPS and  FPS/W are normalized with respect to AMW executing input stationary dataflow (AMW$_{BPCA}$-WS) for ResNet50 at 10 GS/s.} 
  \label{fig11}
\end{figure*}

\begin{figure*}[h]
  \centering
  \includegraphics[width=\linewidth]{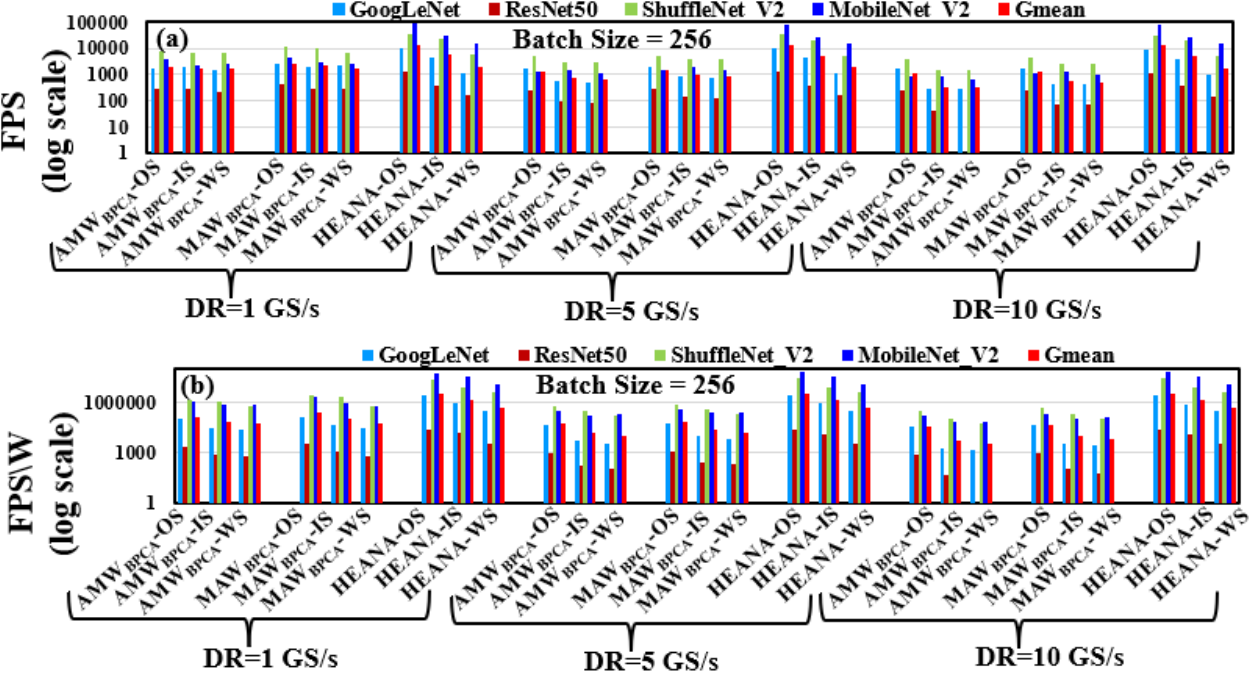}
  \caption{(a) Normalized FPS (log scale), (b) Normalized FPS/W for HEANA versus BPCA-integrated versions of AMW and MAW accelerators with input batch size=256. Results of FPS and  FPS/W are normalized with respect to AMW executing input stationary dataflow (AMW$_{BPCA}$-WS) for ResNet50 at 10 GS/s.} 
  \label{fig12}
\end{figure*}

Fig. \ref{fig11} (a) shows FPS (log scale) results for HEANA versus BPCA integrated AMW$_{BPCA}$ and MAW$_{BPCA}$ accelerators with batch size=1 at different DRs, normalized to AMW-WS for ResNet50 at 10 GS/s. Our accelerator HEANA on gmean across four CNNs, outperforms AMW$_{BPCA}$ and MAW$_{BPCA}$ for \textit{IS}, \textit{OS}, and \textit{WS} dataflows across all datarates. At 1 GS/s, HEANA-OS on gmean achieves up to 6.3$\times$ and 4.6$\times$ better FPS than AMW$_{BPCA}$ and MAW$_{BPCA}$, respectively, across all the dataflows.
At higher datarates such as 5 GS/s and 10 GS/s, our HEANA-OS achieves up to 8$\times$ and 9$\times$ better FPS than AMW$_{BPCA}$ and MAW$_{BPCA}$, across all dataflows. We can observe that the throughput gap between HEANA and BPCA-integrated AMW and MAW architectures decreases, with the integration of BPCA, AMW$_{BPCA}$ and MAW$_{BPCA}$ architectures leverage in-situ temporal accumulations at BPCA reducing the latency corresponding to the reduction of \textit{psum}s. Therefore, AMW$_{BPCA}$ and MAW$_{BPCA}$ achieve better FPS than baseline AMW and MAW. With the integration of BPCA, the throughput order of dataflows remains unchanged in AMW$_{BPCA}$ and MAW$_{BPCA}$. Although \textit{IS} and \textit{WS} dataflow exploit BPCA in-situ accumulation, they still need frequent switching of capacitors at BPCA to reduce \textit{psums} corresponding to various output pixels. In contrast, \textit{OS} dataflow performs temporal accumulation without requiring capacitor switching at BPCA (refer Section \ref{section4}), thus giving better throughput than \textit{IS} and \textit{WS} dataflow.

Fig. \ref{fig11}(b) shows FPS/W (log scale) results for HEANA versus BPCA integrated AMW$_{BPCA}$ and MAW$_{BPCA}$ accelerators with batch size=1 at different DRs, normalized to AMW-WS for ResNet50 at 10 GS/s. Our accelerator HEANA on gmean achieves better FPS/W across four CNNs, outperforms AMW$_{BPCA}$ and MAW$_{BPCA}$ for IS, OS, and WS dataflows across all datarates. At 1 GS/s, HEANA-OS on gmean achieves up to 5.4$\times$ and 3.6$\times$ better FPS/W than AMW$_{BPCA}$ and MAW$_{BPCA}$, respectively, across all the dataflows. At higher datarates such as 5 GS/s and 10 GS/s, our HEANA-OS achieves up to 35$\times$ and 26$\times$ better FPS than AMW$_{BPCA}$ and MAW$_{BPCA}$, across all dataflows. As mentioned before, HEANA's TAOMs reduce the static power consumption of MRRs by using a single MRR design which significantly reduces the power consumption.  
 
Fig. \ref{fig12}(a) and Fig. \ref{fig12}(b) show FPS (log scale) and FPS/W (log scale) results for HEANA versus BPCA-integrated versions of AMW$_{BPCA}$ and MAW$_{BPCA}$ accelerators with batch size=256 at different DRs, normalized to AMW-WS for ResNet50 at 10 GS/s. HEANA-OS on gmean achieves up to 23$\times$ better FPS than other architectures across all datarates. Similarly, HEANA also achieves higher FPS/W for batch size=256. HEANA-OS on gmean achieves up to 92$\times$ better FPS/W than other architectures across all the datarates. We found that the impact of dataflow choice is similar to that of batch size=1.

The area efficiency values (FPS/W/mm$^2$) for each accelerator across various CNNs are similar to the energy efficiency (FPS/W) values for area proportional analysis (when the area of all the accelerators is matched to the area of HEANA). Therefore, we have not reported area efficiency results. Overall, HEANA significantly improves throughput (FPS) and energy efficiency (FPS/W) across various data rates and dataflows compared to the tested analog optical accelerators.

\subsection{Inference Accuracy Results}
As discussed in Section \ref{sec321}, HEANA provides several advantages compared
to prior analog accelerators, but TAOMs of HEANA incur some inaccuracies in dot product computation (see Section \ref{sec323}) due to the possible calibration errors and analog noise. To evaluate the impact of these inaccuracies on CNN inference, we evaluated the inference accuracy of CNN models on HEANA. We integrated our custom simulator with ML-framework PyTorch \cite{pytorch} and incorporated a mean absolute error model of TAOM-based multiplication operation (see Section \ref{sec323}) in the simulator. Then, using this simulator, we performed inference on ImageNet validation dataset \cite{2022ActiveloopHub} (50k images and 1k classes). Table \ref{accuracy} reports the Top 1 and Top 5 inference accuracy across various 8-bit quantized CNN models on the analog architectures and HEANA. Our HEANA accelerator results in Top 1 and Top 5 errors of 0.1\% and 0.1\%, respectively, on average across CNN models. Additionally, we also evaluated the accuracy drop experienced by different CNN models at 4-bit quantized weights and activations by employing the quantization techniques from \cite{robbanner, lqnets, profit}. Table \ref{accuracy2} reports the Top 1 accuracy drop for these models at 4-bit quantization.  Please note
that these accuracy values reported in the Table \ref{accuracy2} also account for the analog optical
computing errors. Our HEANA accelerator's excellent gains in FPS and FPS/W make this minor drop in CNN inference accuracy tolerable. 

\begin{table}[]
\centering
\caption{Top-1 and Top-5 inference accuracy comparison of HEANA versus MAW for 8-bit quantized CNNs \{GoogleNet (GNet), ResNet50 (RNet50), MobileNet\_V2 (MNet\_V2), ShuffleNet\_V2 (SNet\_V2)\} and ImageNet dataset \cite{imagenet}.}
\label{accuracy}
\begin{tabular}{|c|c|c|c|c|c|}
\hline
\begin{tabular}[c]{@{}c@{}}\textbf{HEANA}\\\textbf{ ACCURACY} \\ \textbf{DROP} (\%)\end{tabular} &\begin{tabular}[c]{@{}c@{}}\textbf{GNet}\\ \textbf{\cite{googlenet}}\end{tabular} & \begin{tabular}[c]{@{}c@{}}\textbf{RNet}\\ \textbf{\cite{resnet}}\end{tabular} &  \begin{tabular}[c]{@{}c@{}}\textbf{MNet\_V2}\\ \textbf{\cite{mobilenetv2}}\end{tabular} & \begin{tabular}[c]{@{}c@{}}\textbf{SNet\_V2}\\ \textbf{\cite{shufflenet}}\end{tabular} & \begin{tabular}[c]{@{}c@{}}\textbf{Gmean} \end{tabular} \\ \hline
\textbf{TOP-1}                                                         & 0                & 0               &   0.1                  & 0.1                   & 0.1\\ \hline
\textbf{TOP-5}                                                         & 0.1                & 0.1               & 0.1                    & 0.1                     & 0.1\\ \hline
\end{tabular}
\end{table}

\begin{table}[]
\centering
\caption{Top-1 inference accuracy of HEANA for 4-bit quantized CNNs \{GoogleNet (GNet), ResNet50 (RNet50), MobileNet\_V2 (MNet\_V2), ShuffleNet\_V2 (SNet\_V2)\} and ImageNet dataset \cite{imagenet}}.
\label{accuracy2}
\begin{tabular}{|c|c|c|c|c|c|}
\hline
\begin{tabular}[c]{@{}c@{}}\textbf{HEANA}\\\textbf{ ACCURACY} \\ \textbf{DROP} (\%)\end{tabular} &\begin{tabular}[c]{@{}c@{}}\textbf{GNet}\\ \textbf{\cite{googlenet}}\end{tabular} & \begin{tabular}[c]{@{}c@{}}\textbf{RNet}\\ \textbf{\cite{resnet}}\end{tabular} &  \begin{tabular}[c]{@{}c@{}}\textbf{MNet\_V2}\\ \textbf{\cite{mobilenetv2}}\end{tabular} & \begin{tabular}[c]{@{}c@{}}\textbf{SNet\_V2}\\ \textbf{\cite{shufflenet}}\end{tabular} & \begin{tabular}[c]{@{}c@{}}\textbf{Gmean} \end{tabular} \\ \hline
\textbf{TOP-1}                                                         
& 1.5                &    2.3            &   0.5                  & 1                   & 1.1 \\ \hline
\end{tabular}
\end{table}

\section{Discussion}\label{625}
HEANA presents several innovations at the circuit level and architecture level to mitigate the shortcomings of prior optical CNN accelerators. At the circuit level, HEANA employs a novel time-amplitude analog optical modulator (TAOM) that generates the product of one input value and weight value as a pulse-width-amplitude-modulated (PWAM) symbol whose optical energy is proportional to the product result. TAOM achieves this by employing a single microring as opposed to the dual microrings used by prior  works\cite{deapcnn,crosslight,holylight}. By using a single microring for each multiplication, HEANA reduces the needed feedback control units to half, which decreases the static power consumption and improves the energy efficiency. Use of single microring per multiplication also reduces the overall optical losses and power penalties, furthering the energy efficiency advantage. In addition, HEANA also employs a balanced photo-charge accumulator (BPCA) which leverages the capabilities of a balanced photodetector to perform in-situ spatio-temporal accumulations. As discussed in Section \ref{section4}, BPCA improves power consumption by reducing the analog-to-digital conversions and buffer accesses, and by completely eliminating the use of a reduction network for accumulations. With all these cumulative advantages HEANA gains at least 32$\times$ better energy efficiency than prior optical accelerators. At the architectural level, HEANA presents a spectrally hitless architecture that mitigates inter-modulation crosstalk which reduces the power penalty, increasing the available power budget. The saved power budget allows HEANA to improve the DPU size from 44$\times$44 to 83$\times$83 at 4-bit precision, this improves spatial parallelism and decreases the number of \textit{psum} reductions required. This in turn significantly decreases the latency corresponding to the analog-to-digital conversions and \textit{psum} reductions to complete the computations by reducing the required analog-to-digital conversions and \textit{psum} reductions. These benefits allow HEANA to achieve at least 23$\times$ better throughput than prior optical accelerators. Thus, due to these cross-layer innovations, HEANA achieves superior performance compared to prior optical CNN accelerators. Furthermore, HEANA has the potential to outperform ReRAM crossbars-based CNN accelerators and other analog photonic accelerators from prior work due to these two advantages: (1) at least an order of magnitude shorter latency of partial dot-product result generation compared to ReRAM crossbars (ReRAM crossbars have about 100ns latency \cite{rramsurvery}, whereas HEANA can achieve sub-nanosecond-scale latency) and (2) significant energy reduction and dataflow flexibility enhancement in HEANA compared to prior photonic accelerators due to HEANA's extended analog operation for partial result accumulation.

\section{Conclusions}

In this paper, we presented HEANA, a novel hybrid time-amplitude analog optical GEMM accelerator. The dot product element of our HEANA employs a spectrally hitless array of novel hybrid time-amplitude analog optical modulators (TAOM) that perform multiplication operations, combined with a balanced photo-charge accumulator (BPCA) that performs spatio-temporal accumulation of the individual multiplication results from TAOMs. Each TAOM utilizes a single active microring resonator (MRR) to perform multiplication operations, allowing for the electro-optic tuning of both input and weight values. This enables the flexibility of executing various CNN dataflows namely input stationary, output stationary, and weight stationary dataflows. Moreover, The spectrally hitless arrangement of TAOMs in HEANA eliminates spectral-interference and various crosstalk effects encountered by current MRR-enabled analog photonic accelerators. In addition, the single MRR implementation of TAOMs substantially reduces the area consumption and the insertion losses in HEANA, thereby increasing its energy efficiency. Also, the BPCA circuit in HEANA enables in-situ accumulation of a large number of partial sums, thereby reducing the overall latency and energy consumption of CNN processing. 

We performed detailed modeling and characterization of our TAOM unit using photonics foundry-validated tools from ANSYS/Lumerical. Here, we performed time-domain simulations from which we evaluated the performance of our TAOM in terms of accuracy and precision. We have also performed a scalability analysis of our HEANA’s dot product units, to determine their achievable maximum size and supported bit-precision. Finally, we evaluated HEANA for input stationary, output stationary, and weight stationary dataflows at the system-level, and compared its performance with two well-known photonic CNN accelerators from prior works. Our system-level evaluation results for four CNN models show that HEANA provides improvements of up to 25$\times$ and 32$\times$ in throughput, and energy efficiency, respectively, compared to two prior optical analog accelerators AMM and MAM, with Top-1 accuracy drop of only up to 0.1\%.

\section*{Acknowledgments}
 We would like to acknowledge the National Science Foundation (NSF) as this research was supported by NSF under grant CNS-2139167.

\bibliographystyle{ACM-Reference-Format}
\bibliography{reference}

\end{document}